\documentclass[preprint2]{aastex}

\bibpunct{(}{)}{;}{a}{}{,}
\newcommand{\FIG}[1]{#1}

\usepackage{txfonts}
\usepackage{graphicx}
\usepackage{natbib}

\slugcomment{ }

\shorttitle{Relativistic two-component jets}
\shortauthors{Z. Meliani}

\begin{document}
\title{Decelerating relativistic two-component jets}

\author{Z. Meliani
           \altaffilmark{1}, R. Keppens \altaffilmark{1,2,3}}

\altaffiltext{1}{Centre for Plasma Astrophysics, K.U.Leuven, Belgium}
\altaffiltext{2}{FOM-Institute for Plasma Physics Rijnhuizen, Nieuwegein, \\
                 The Netherlands}
\altaffiltext{3}{Astronomical Institute, Utrecht University, The Netherlands}
 \email{Zakaria.Meliani@wis.kuleuven.be, Rony.Keppens@wis.kuleuven.be}

\date{Received ... / accepted ...}
\begin{abstract}

Transverse stratification is a common intrinsic feature of astrophysical jets. 
There is growing evidence that jets in radio galaxies consist of a fast low density outflow at the jet 
axis, surrounded by a slower, denser, extended jet. The inner and outer jet components then have a 
different origin and launching mechanism, making their  effective inertia, magnetization, associated energy flux and 
angular momentum content different as well. Their interface will develop differential rotation, where 
disruptions may occur. We here investigate the stability of rotating, two-component relativistic outflows typical for jets in radio galaxies. For this purpose, we parametrically explore the long term evolution  of a transverse cross-section of 
radially stratified jets numerically, extending our previous study where a single, purely hydrodynamic evolution was considered. 
 We include cases with poloidally magnetized jet components, covering hydro and magnetohydrodynamic models.
With grid-adaptive relativistic magnetohydrodynamic simulations, augmented with approximate linear stability analysis, we revisit the interaction 
between the two jet components. We study the influence of dynamically important  poloidal magnetic fields, with varying 
contributions of the inner component jet to the total kinetic energy flux of the jet, on 
their non-linear azimuthal stability.
We demonstrate that two-component jets with high kinetic energy flux, and an inner jet  effective inertia which is higher 
than the outer jet effective inertia are subject to the development of a relativistically enhanced, rotation-induced Rayleigh-Taylor type instability. This instability plays a major role in decelerating the inner jet and the overall jet decollimation.
This novel deceleration scenario can partly explain the radio source dichotomy, relating it directly to the 
efficiency of the central engine in launching the inner jet component. The FRII/FRI transition could then 
occur when the relative kinetic energy flux of the inner to the outer jet grows beyond a certain treshold.
\end{abstract}
\keywords{ISM: jets and outflows -- Galaxies: jets -- methods: numerical, relativity}

 
\section{Introduction}
There is strong observational and theoretical evidence that magnetic fields play 
a crucial role in the acceleration and the collimation of extragalactic jets. 
Most Active Galactic Nuclei (AGN) jet formation scenarios involve magnetic fields threading a rotating 
black hole (in the ergosphere) and its accretion disk, thereby removing from them angular momentum, 
allowing the central black hole to accrete. At least close to the jet launching region, the jet rotation 
profile persists, reflecting its (general relativistic and/or magneto-rotational) origin. 
Thus both ingredients, magnetic fields and 
rotation, are very important in jet formation, as well as in jet propagation and stability. 

Moreover, detailed astrophysical jet observations point out that relativistic jets are structured, 
in a direction perpendicular to the jet axis, typically consisting of a fast spine and slower outer flow. In 
the case of AGNs, this jet structuring plays an important role in explaining the morphology of the jet 
high energy radiation 
\citep{Ghisellinietal05,Hardcastle06, Jesteretal06,Jesteretal07,Siemiginowskaetal07,Kataokaetal08}, 
with sometimes clear evidence for a very fast, light inner jet and a heavy slow outer 
outflow \citep{Girolettietal04}. Furthermore, observations of the TeV BL Lacertae objects show 
brightenings and rapid variability in their TeV emission. This variation in their high 
energy emission implies high Lorentz factor flows occuring at smaller scale, suggesting ultra-relativistic 
bulk motion of the (inner) jet.
At the same time, complementary (radio) observations with VLBI of the pc-scale jet structure indicate a broad,
`slowly' (albeit relativistic) moving outflow. In combination, this clearly suggests the presence of a two component jet 
morphology \citep{Ghisellinietal05}. 
Two-component jet structure has also been proposed
in more theoretical work, addressing the physics of jet launching, collimation and propagation 
mechanisms \citep{BogovalovTsinganos01, Soletal89, Meier03}.

While our jet dynamics computations will be representative for AGN jet conditions, radially structured
jet flows are now known to exist in virtually all astrophysical jet outflows.
Transversely structured, ultra-relativistic jet-like outflow has been proposed in the context of 
Gamma Ray Bursts \citep{Racusinetal08}, to explain the break observed in their afterglow light curve. 
In the case of stellar outflows, recent observations of some 
T Tauri jets \citep{Bacciottietal00, Guntheretal09} also suggest a fast 
inner outflow bounded by a slow outer outflow. 
In these young stellar objects, a clear signature of jet rotation around the symmetry axis was 
detected \citep{Bacciottietal02, Woitasetal05,Coffeyetal04,Coffeyetal07}, fully supporting scenarios of 
magneto-centrifugal jet launch and acceleration. Theoretical models of two-component jets in classical 
T Tauri \citep{BogovalovTsinganos01,Melianietal06,Cranmer08,Fendt09} then postulate that the inner outflow is 
turbulent and pressure driven, associated with the young star wind. The inner jet then has a small opening 
angle, as it is collimated by the outer jet, which is in turn magneto-centrifugally driven from the surrounding disk.
The outer disk wind then carries most of the mass loss in the jet. Various 
authors \citep{Melianietal06, Fendt09} have demonstrated  using axisymmetric MHD simulations that the outer outflow is self collimated by 
its intrinsic magnetic field, and that the turbulent inner outflow gets collimated by the outer jet. 
Furthermore, \cite{Matsakos08} investigated the topological stability of two-component outflows for young stellar
objects, performing extensive numerical simulations to determine whether analytic self-similar models
demonstrate robustness in axisymmetric conditions. 

 Also for relativistic jet simulations, axisymmetry assumptions are often adopted, excluding the development of all 
non-axisymmetric perturbations. These can address details of how helical field configurations (naturally expected from
magneto-centrifugal launch mechanisms) effectively may transport their helicity down the jet beam \citep{Keppensetal08}, 
with magnetically aided reacceleration by field compression across internal cross-shocks. While \cite{Keppensetal08} concentrated
on kinetic energy dominated jets, initially Poynting flux dominated jets were simulated by \cite{Komissarovetal07} in axisymmetric
relativistic MHD, finding that the transition to a 
matter-dominated jet regime occurs very close to the central engine (within $0.01 {\rm pc}$). Our model computations will therefore assume kinetic energy dominated jets.

As far as the magnetic field topology is concerned, we will restrict ourselves in this paper to purely poloidally magnetized
jet components. Our two-component jet model determining our initial conditions can actually allow for helical fields, 
as explained in Sect.~\ref{s-model} (we include this more general case here, for future reference in follow-up studies).
 As indicated before, during the first acceleration phase of AGN jets, magneto-centrifugal mechanisms play an important role 
and a helical or even strongly toroidal magnetic field is likely 
produced \citep{Fendt97,Melianietal06a, McKinney&Blandford09,Komissarovetal07}. \cite{McKinney&Blandford09} present 3D general relativistic MHD simulations for rapidly rotating black holes producing jets with strong toroidal fields. They find a prominent role of the accreted magnetic field geometry for achieving `stable' jets. 
Helical or strongly toroidal field topologies can be subject to current-driven kink
instabilities \citep{Begelman98}, with $m=1$ toroidal modes that helically displace the jet axis. This requires full 3D numerical
simulations, such as performed by \cite{BatyKep02} in non-relativistic MHD, or addressed by \cite{Mizunoetal09} in relativistic 
MHD for a static force-free equilibrium.
Dispersion relations for non-axisymmetric modes and $m=1$ kinks in particular for relativistic MHD were analysed
by \cite{Begelman98} for purely toroidal fields, and electromagnetically dominated, force-free jets were analysed spectrally by \cite{Istomin&Pariev96} and more recently by~\cite{Narayanetal09}.

\begin{figure}[h]

\begin{center}

{\resizebox{0.95\columnwidth}{6cm}{\includegraphics{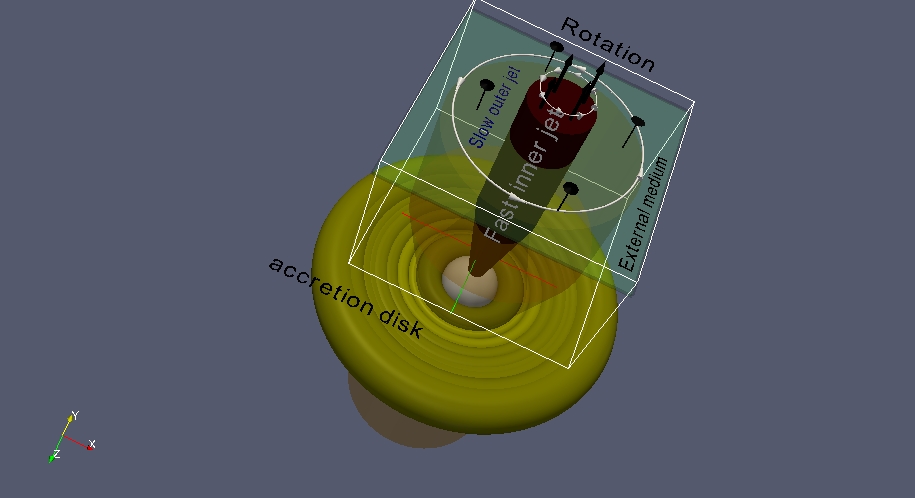}}}

\end{center}
\caption{3D schematic view of the overall AGN disk-jet configuration (indicating the accretion disk and the two-component jet). 
We model the jet evolution in the transverse plane.}\label{Fig1}
\end{figure}

 In our work, we will restrict attention to 2.5D
scenarios with the somewhat unusual assumption of translational invariance along the jet axis. The overall configuration is schematically indicated in Fig.~\ref{Fig1}, and we simulate a transverse cross-section of the jet at sufficient distance from the two-component jet source, where all three velocity components (axial, azimuthal and radial) are included, but their variation along the jet axis is ignored.
Our aim is to investigate all non-axisymmetric instabilities, primarily induced by the (sheared) rotation. This approximation is 
valid because in the poloidal direction, the flow is supersonic with high Lorentz factor, and then the growth rate of poloidal 
instabilities is expected to be low. On the other hand, the rotation is subsonic, facilitating the growth of toroidal 
instabilities.
We then address jet stability in a cross section of a rotating two-component jet, initially collimated by thermal pressure 
and/or poloidal magnetic field. 
Consecutive snapshots of the cross-sectional evolution can be interpreted as mimicking the jet flow conditions at 
increasing distance from the source.  Note that this particular assumption allows us to follow both axisymmetric and
all non-axisymmetric (including $m=1$) mode development, but does exclude helical mode axis displacement typical for kink modes.
Our model therefore mimics jet evolution adequately as long as the radial axis displacement is smaller than the axial wavelength
associated with possible $m=1$ kinks. Our assumption does also neglect conical jet expansion, assuming cylindrical propagation. This is justified given the low observed values for jet opening angles. 

 Since we defer the study of toroidal and helical magnetic field configurations in 2.5D and 3D to later work,  
we start off with numerically investigating
the influence of a purely poloidal magnetic field on the stability of rotating, two-component relativistic jets.
 Our work complements the studies looking into kink development, by putting the emphasis on the azimuthal variation and on the effect of the two-component jet stratification.
As far as a purely poloidal magnetic field topology is concerned, the work by \cite{Spruitetal97} suggests that jets should 
be collimated by poloidal magnetic field pressure, rather than by toroidal magnetic field, as  
toroidal jet magnetic fields can introduce kink instability (but only slow mode growth was found for force-free jets by~\cite{Narayanetal09}). 
To justify purely toroidal fields, we can argue that the toroidal magnetic field in the jet acceleration phase can be gradually dissipated involving reconnection, a mechanism which in turn contributes to (axial) jet  acceleration \citep{Spruit08, Melianietal06}. Indeed, during the acceleration phase 
 a fraction of the Poynting flux (angular momentum) carried by the magnetic field is converted to kinetic energy by internal dissipation/reconnection of the toroidal magnetic field~\citep{Spruit08, Melianietal06}. In accord with this mechanism, we model the region where the eventual jet rotation is low.


In this paper, we analyse five cases in detail, to determine the effects of  differing poloidal magnetic field configurations in the
two-component structure on its long-term stability, and on the overall deceleration of the jet as it propagates away from the central
source regions. 
 The role of poloidal magnetic field is in these cases most prominent in its added effect on total pressure and effective fluid inertia, and this is shown to play a prominent role in the two-component jet stability.

\section{Two-component jet model}\label{s-model}

We set up a relativistic magnetohydrodynamic model of a two-component jet, which elaborates on the earlier
model described in \cite{Meliani&Keppen07b}. The jet model uses
as its basic characterising parameter a total jet kinetic luminosity flux typical for powerful radio loud 
galaxies, namely $L_{\rm jet, Kin} = 10^{46} {\rm ergs/s}$ \citep{Rawlings&Saunders91, Tavecchioetal04}.
Secondly, the outer radius of the two-component jet is taken to be $R_{\rm out} \sim 0.1 {\rm pc}$.
This is a value directly inferred from observed values for M87 \citep{Birettaetal02}, which
has a known jet opening angle of $6^{\circ}$ at $1 {\rm pc}$ from the jet source. For the less constrained
inner jet radius, we adopt the initial value $R_{\rm in} =  R_{\rm out}/3$. The remainder of the $t=0$
condition is then characterized by the velocity profiles in both inner and outer jet components and by the
magnetic configuration in the two-component jet. 

\subsection{Initial flow profiles and Rayleigh criterion}\label{Initial_cond}

For the initial flow condition, we adopt a uniform outflow velocity $V_{z}$ along the jet axis in each 
jet component, with $V_{\rm z, out} = 0.9428$ (hence $\gamma_{\rm z, out} \sim 3$,  since speeds are 
normalised to the light speed) for the outer, slow jet located within $R_{\rm in} < R < R_{\rm out}$. 
This represents a typical outflow value for relativistic jets at parsec scale from the source. The inner
jet has a much faster outflow speed, which we set to $V_{\rm z, in} =0.99939$, with corresponding Lorentz 
factor $\gamma_{\rm z, in} \sim 30$. The difference in Lorentz factor
by an order of magnitude in the inner versus outer jet
layers is representative (though somewhat exaggerated) for differences inferred for the BL Lac object
Markarian 501 between its central spine and surrounding (shear) layer 
jet, as described by \cite{Girolettietal04}. However, it remains lower than the 
Lorentz factor $\gamma\sim 50$ suggested for TeV blazar PKS 2155-304 \citep{Giroletti&Tavecchio08}.

Two-component jet models also commonly assume that the spin of the inner beam is higher than the spin of the outer jet. The physical argument is that the inner beam is extracting angular momentum from the inner parts 
of the accretion disk and from the black hole itself, both implying fast rotation.
Therefore, we adopt different initial rotation profiles for inner versus outer jet, both radially self-similar.
Although we will in our numerical results below only adopt purely poloidal magnetic 
field configurations, i.e. we will have $\vec{B}=B_z(R,\varphi; t)\vec{e}_z$, one can generalize our initial equilibrium
configuration to allow for both toroidal speeds and toroidal magnetic fields, sharing the same $R$-dependence. 
In that general case, one takes in practice
\begin{equation}
V_{\varphi}=\left\{
\begin{array}{ccr}
v_{\varphi, \rm in} \left(\frac{R}{R_{\rm in}}\right)^{a_{\rm in}/2}& & R \le R_{\rm in}\,,\\
v_{\varphi, \rm out} \left(\frac{R}{R_{\rm in}}\right)^{a_{\rm out}/2}& & R_{\rm in}<R<R_{\rm out} \,,
\end{array}
\right.
\label{q-vphi}
\end{equation}
\begin{equation}
B_{\varphi}=\left\{
\begin{array}{ccr}
b_{\varphi, \rm in} \left(\frac{R}{R_{\rm in}}\right)^{a_{\rm in}/2}& & R \le R_{\rm in}\,,\\
b_{\varphi, \rm out} \left(\frac{R}{R_{\rm in}}\right)^{a_{\rm out}/2}& & R_{\rm in}<R<R_{\rm out} \,.
\end{array}
\right.
\label{q-bphi}
\end{equation}
 where the $\vec{V}=V_{\varphi}\vec{e}_\varphi+V_{\rm z}\vec{e}_z$ is the 3-velocity, and similarly for the magnetic field.
At $t=0$, we then have a discontinuity in the toroidal velocity at the boundary between the two
jets when fixing $v_{\varphi,\rm in} = 0.01$ and $v_{\varphi, \rm out}=0.001$.
A faster rotating inner component is consistent with a small expansion of the inner streamlines, which are
confined by the outer outflow. The fast rotating inner jet effectively extracts the angular 
momentum from the central region and carries it away with a very small mass flux. In fact, our models typically
have the inner jet extracting order $10\%$ of the total angular momentum associated with the two-component 
jet. We fix the exponent $a_{\rm in}=0.5$ for the inner jet and $a_{\rm out}=-2$ for the outer jet.
We then choose for the outer jet a rotation profile implying a radially constant angular momentum.
This is motivated by the fact that this component is believed to be launched from an
accretion disk, where the outer streamlines in the jet expand faster than the 
inner streamlines. 

The rotation profile of the inner jet then obeys the relativistic equivalent of
the Rayleigh criterion for stability, namely that the angular momentum flux given by
\begin{equation}
I=\gamma \frac{\left(\rho+\frac{\Gamma}{\Gamma-1}p\right)}{\rho}\,V_{\varphi}\,R-\frac{B_{\rm p}}{\gamma\,\rho\,V_{\rm p}}\,R\,B_{\varphi}\,,
\label{q-am}
\end{equation}
increases with $R$ \citep{Melianietal06a}. This expression uses proper density and pressure $\rho$, $p$, assumes a polytropic index $\Gamma$ appearing in
a simple polytropic equation of state (our actual numerical simulations will in fact relax this
assumption and use a full Synge-type equation of state), and writes $V_p$, $B_p$ for poloidal three-velocity and magnetic field strengths. 
For the chosen radial prescriptions for $V_\varphi$ given by Eq.~(\ref{q-vphi}), we have for the inner jet
\begin{equation}
\frac{{\rm d} |I|}{{\rm d} R}\propto \left(a_{\rm in}/2+1\right)>0\,,
\end{equation}
making the inner jet centrifugally stable. The outer outflow on the other hand is marginally 
stable, following the same argument. In the numerical simulations further in this paper, we
neglect toroidal field $B_\varphi$, and then
the interface between the two components does not verify the 
Rayleigh criterion and this shear flow interface is unstable from the start, since the angular momentum 
at the interface decreases when  
\begin{equation}
\gamma_{\rm out}\frac{\left(\rho_{\rm out}+\frac{\Gamma_{\rm out}}{\Gamma_{\rm out}-1}p_{\rm out}\right)}{\rho_{\rm out}}
v_{\varphi, \rm out}<\gamma_{\rm in}\frac{\left(\rho_{\rm in}+\frac{\Gamma_{\rm in}}{\Gamma_{\rm in}-1}p_{\rm in}\right)}{\rho_{\rm in}} v_{\varphi,\rm in}.
\end{equation}
This means that in all cases we will simulate, the initial shear flow interface at $R=R_{{\rm in}}$ will develop
small-scale instabilities, and we intend to address the ultimate non-linear stability of such initial
two-component structure. The central question investigated then is whether the dynamics will lead to complete destruction of
the initial two-component nature of the outflow at further distances from the source. Related to this question, we
will quantify the potential deceleration of the central fast jet by mixing processes, induced by the
nonlinear evolution of the two-component jet.

\subsection{Initial densities and magnetic configuration}

The densities and the poloidal magnetic field $B_{\rm z}$ are assumed to be constant within the
three different regions, namely throughout inner jet beam, outer jet, and external region. 
The external medium is hot and rarified and has a small reference number density.  The typical ISM number density value of $1 {\rm cm}^{-3}$ is in the computation used as a scaling value. We arbitrarily set $\rho_{\rm med}=m_p n_{\rm med}$ with $n_{\rm med}=10^{-2} {\rm cm}^{-3}$, and adopt $c=1$ and a unit of length of $1{\rm pc}$. This sets the unit of pressure, and a corresponding scaling value for the magnetic 
field is then $0.1375\,{\rm Gauss}$ (as mass is normalized to proton mass $m_p$).

The jet environment represents a relativistically hot, rarified, static external medium.
This is consistent with the fact that at the parsec scale, the jet head has previously shocked the jet 
surroundings, so that the jet itself gets embedded in a dilute, hot medium.
Assuming a number density for this external medium of $n_{\rm med}=10^{-2} {\rm cm}^{-3}$ is appropriate, since this
is near values obtained in numerical simulations of jet propagation and are then typical
for the jet cocoon surroundings \citep{Melianietal08}. Values for the jet component proper densities are 
estimated from the kinetic luminosity flux and its division over inner and outer jet components. 
Requiring that the outer jet carries a case-specific fraction of the total kinetic luminosity flux by
\begin{equation}
f_{\rm out} L_{\rm jet, Kin} =  \left(\gamma_{\rm out}\,h_{\rm out}-
1\right)\rho_{\rm out}\gamma_{\rm out}\pi
\,(R_{\rm out}^2-R_{\rm in}^2) V_{\rm out}\,, \label{kinjet}
\end{equation}
we can deduce the density $\rho_{\rm out}$, if we in addition prescribe the (outer) jet Lorentz factor 
and make the initial approximation that the thermal energy in the outer
jet is negligible compared to its mass energy (as valid in all cases studied). 
 In Eq.~(\ref{kinjet}), $h_{\rm out}$ is the specific enthalpy in the outer jet, and its expression depends on the equation of state.
When we consider that the inner component carries only a small fraction $f_{\rm in}= 1\%$ of the total 
kinetic luminosity flux, such that the outer jet carries the remaining $f_{\rm out} = 99\%$, we find
that the resulting density is $\rho_{\rm out}\sim 119.94\times 10^{2}\rho_{\rm med}$.
A similar argument for the inner, light jet, works out to 
fix $\rho_{\rm in}\sim 6.92 \rho_{\rm med}$. 

While in all cases we investigate here numerically, we assume these initial, piecewise constant, densities for the inner and outer jet,
the actual kinetic luminosity carried by each component will change from one case to another. This relates to
our prescription of the pressure and magnetic variation in the jet components, and we will simulate five cases 
with varying thermal pressure contribution at the jet axis.
In four cases (A), (B2), (C) and (D), the total pressure at the jet axis ends up similar, while case (B1) has a 
lower total pressure value.
By varying the relative contribution of thermal to magnetic pressure in the jet components, we
aim to analyze the effect on overall jet stability of varying magnetic field and kinetic energy contributions 
to the kinetic luminosity in each component. In all five cases studied, we set the pressure of
the external medium to ensure total pressure balance with the outer jet. 
Also, in the  four cases (A), (B1), (C), (D), we assume that the thermal energy in the inner (hot) component is higher than its mass energy. Only in the case (B2), the thermal energy is lower than the mass energy.
They differ in the following sense: in the first case (A), both components are non-magnetized and the pressure at the jet axis is $p_0=2.3$  in our simulation units.
This purely hydrodynamic case differs from the hydro case already studied in~\cite{Meliani&Keppen07b}, 
as now the inner jet has an even higher Lorentz factor, a slightly different inner rotation profile, and an on axis pressure 
value which makes the entire inner jet relativistically hot. In the cases (B1) and (B2), only the inner jet is magnetized and we set it to a constant 
value $B_{\rm z, in}= \sqrt{0.01 \gamma^{2}_{\rm in}\rho_{\rm in}}\sim 0.789$ (making the jet parameter
$\sigma=B_{\rm z}^2/(\gamma^2 \rho)$ expressing magnetic to rest-mass energy of order $0.01$, as for kinetic energy dominated jets) in the case (B1) and $B_{\rm z, in}\sim 2.28$ in the case (B2). In the case (B1), the thermal pressure on axis is 
set to $p_0=0.1$ and in the case (B2) the thermal pressure on axis is $p_0=10^{-4}$. In the case (C), only the outer jet is magnetized with $B_{\rm z, out}=\sqrt{0.005 \gamma^{2}_{\rm out}\rho_{\rm out}}\sim 2.323$, 
and we adopt the same order on axis inner pressure as in (A), namely $p_0\approx 2.3$. 
In case (D), both inner and outer jet are magnetized, with $B_{\rm z, in}\sim 0.789$ and $B_{\rm z, out}\sim 2.323$. We then take the 
pressure at the jet axis  $p_0\approx 2$. In all magnetized cases, the magnetic field strenght is set according to the observations at parsec-scale \citep{OSullivan&Gabuzda09}.
Clearly, in all four cases (A), (B1), (C) and (D), 
the thermal energy in the inner component dominates, only in the case (B2) the thermal energy is lower than the 
mass energy.
The pressure profile through inner and outer jet is taken from transverse equilibrium conditions, discussed next.

\subsection{Two-component jet pressure profiles}

Since we assume an initial near steady-state axisymmetric two component jet, the actual pressure variation is deduced from 
transverse equilibrium between pressure gradient, centrifugal force, and Lorentz force. This is expressed by the steady-state equation \citep{Melianietal06a, Appl&Camenzind93, Heyvaerts&Norman03}
\begin{equation}\label{eqeuler}
{\gamma \rho \vec{V} \cdot \nabla} (\gamma h \vec{V}) = -\nabla{p}+\rho_{e}\vec{E}+\vec{J_{\rm e}}\wedge\vec{B}\,.
\end{equation}
Here $\rho$ is the proper density of particles, $p$ and $h$ are the pressure and enthalpy per particle, respectively, $\vec{V}$ is the fluid three-velocity, $\gamma$ the Lorentz factor, and ($\vec{E}, \vec{B}$) denote
the electromagnetic fields. We have $\rho_{\rm e}=\nabla{\cdot\vec{E}}$ and for steady-state conditions
$\vec{J_{\rm e}}=\nabla\wedge\vec{B}$ as the associated charge and current densities, while $\vec{E}=-\vec{V}\wedge\vec{B}$ in ideal relativistic MHD.

If we introduce the  total pressure in the fluid frame,
\begin{equation}\label{pseudoPressure}
\psi=p+\frac{B^2-E^2}{2}=p+\frac{B_{z}^2+B_{\varphi}^2}{2}-\frac{\left(B_{\varphi} V_{z}-V_{\varphi} B_{z}\right)^{2}}{2}\,,
\end{equation}
the radial component of Eq.~(\ref{eqeuler}) writes as
\begin{eqnarray}\label{difpseudoPressure}
\frac{{\rm d} \psi}{{\rm d} R}&-&\frac{\Gamma}{\Gamma-1}\frac{\gamma^2 V_{\varphi}^2}{R}\psi=\frac{\gamma^2 V_{\varphi}^2}{R}\left(\rho-\frac{\Gamma}{\Gamma-1}\frac{B_z^2}{2}\right)\nonumber \\
& +&\frac{1}{R}\left(1+\frac{\Gamma \gamma^2 V_{\varphi}^2}{2\left(\Gamma-1\right)}\right)
\left(- B_{\varphi}^2+\left(B_{\varphi} V_{z} - V_{\varphi} B_{z}\right)^2\right) \,.\nonumber \\
& & 
\end{eqnarray}
This has assumed a polytropic equation of state.
We can then determine the radial pressure profile $p(R)$ by solving the first order 
differential equation~(\ref{difpseudoPressure}) and using~(\ref{pseudoPressure}), together with the
self-similar prescriptions from Eq.~(\ref{q-vphi})-(\ref{q-bphi}) to get
\begin{eqnarray}\label{pressure}
p&=&\zeta \left(1-\alpha\,\left(\frac{R}{R_{\rm in}}\right)^{a}\right)^{-\frac{\Gamma}{a\left(\Gamma-1\right)}}
  -\frac{\left(\Gamma-1\right)}{\Gamma}\rho \nonumber\\
 &-&\frac{\left(\Gamma-1\right)\left(a+2\right)}{2\,\alpha\left(a\,\left(\Gamma-1\right)+\Gamma\right)}\left(-b_{\varphi}^2+\left(b_{\varphi} V_{z} - v_{\varphi} B_{z}\right)^2\right)\nonumber\\
&\times&\left(1-\alpha\left(\frac{R}{R_{\rm in}}\right)^{a}\right) \,. 
\end{eqnarray}
In this expression, the constant $a$ is a self-similarity exponent, which we use to set the inner jet exponents $a_{\rm in}$ and the outer $a_{\rm out}$, and the constant $\alpha=\frac{v_{\varphi}^2}{1-V_{\rm z}^2}$, which is
different for inner versus outer jet $\alpha_{\rm in}$, $\alpha_{\rm out}$. 
Also, $\zeta$ is an integration constant, which is to be deduced from boundary conditions. 
For the inner jet, we use it to set the pressure on axis $p_0$ from
\begin{eqnarray}
\zeta_{\rm in} &=& p_0+\frac{\left(\Gamma_{\rm in}-1\right)}{\Gamma_{\rm in}}\rho_{\rm in} \nonumber \\
& & +\frac{\left(\Gamma-1\right)\left(a+2\right)}{2\,\alpha\left(a\,\left(\Gamma-1\right)+\Gamma\right)}\left(-b_{\varphi}^2+\left(b_{\varphi} V_{z} - v_{\varphi} B_{z}\right)^2\right) \,. \nonumber \\
& &
\end{eqnarray}
To fix the constants for the outer jet, we have to match conditions at the interface between 
the two components. At the contact between the two components, Rankine-Hugoniot conditions should hold. 
In terms of effective total pressure~(\ref{pseudoPressure}) obeying Eq.~(\ref{difpseudoPressure}), we
need to match at the interface 
\begin{equation}
\psi_{\rm in}(R_{\rm in})=\psi_{\rm out}(R_{\rm in}) \,.
\end{equation}
This yields for the constant pressure integration in Eq.~(\ref{pressure}) for the outer jet
\begin{eqnarray}
\zeta_{\rm out}&=&
\left[\psi_{\rm in}(R_{\rm in})+\frac{\Gamma_{\rm out}-1}{\Gamma_{\rm out}}\rho_{\rm out}-\frac{B_{z, \rm out}^2}{2}\right]
\left(1-\alpha_{\rm out}\right)^{\frac{\Gamma_{\rm out}}{a_{\rm out}\left(\Gamma_{\rm out}-1\right)}}\nonumber\\
&+&\left[-b_{\varphi, \rm out}^2+\left(b_{\varphi, \rm out} V_{z, \rm out} - v_{\varphi, \rm out} B_{z, \rm out}\right)^2\right]{\left(1-\alpha_{\rm out}\right)^{\frac{\Gamma_{\rm out}}{a_{\rm out}\left(\Gamma_{\rm out}-1\right)}}}\nonumber\\
&\times&\left(\frac{1}{2}
+\frac{\left(1-\alpha_{\rm out}\right)\left(\Gamma_{\rm out}-1\right)\left(a_{\rm out}+2\right)}{2\,\alpha_{\rm out}\left[a_{\rm out}\left(\Gamma_{\rm out}-1\right)+\Gamma_{\rm out}\right]}\right)\,.
\end{eqnarray}
This fixes the pressure variation throughout both components. In the external medium, the thermal pressure is constant and set equal to
the total pressure in the outer jet component. This is actually a slightly different value for cases (A), (B1) and (C), (D); albeit of the same order.
The governing equation of state is taken as a Synge-type relation, also used in \cite{Melianietal04}.
As a result of the above profile prescriptions, the outer jet is relativistically cold with an effective polytropic index $\Gamma_{\rm eff}=5/3$ and the 
inner jet has ultra-relativistic state $\Gamma_{\rm eff}=4/3$. Case (B2) is an MHD variant to the hydro case with more stratified
inner effective polytropic index already simulated in \cite{Meliani&Keppen07b}.
 The matter state between the two components is different (relativistically hot inner and cold outer jet) and we will see that during the time evolution, a shear region with intermediate matter state forms. This variation in the matter state between 
various jet regions makes it vital to use a Synge-EOS to model growing instabilities and resulting turbulence. However, in the 
initial conditions of all cases, the polytropic index is relatively constant throughout each component. This is why we could 
use the polytropic EOS assumption to deduce a {\it near}-equilibrium solution of the initial two-component jet.

\subsection{Dimensionless characterization and numerical setup}

As a result of the initializations described above, the local fast magnetosonic speed in the inner 
jet for case (D) 
and the local sound speed in the inner jet for cases (A) and (C) are of the order of $0.6$ (light speed units), while cases (B1) and (B2) have a higher local fast magnetosonic speed.
In the outer jet, the local fast magnetosonic speed of cases (C), (D) and the local sound speed in case (A) and (B2) are of the order of $0.2$, while it is $0.07$ in the case (B1). 
Both inner and outer jets are kinetically dominated. 
The inner jet has an effective relativistic Mach number $M_{\rm fast}=\frac{\gamma\,V_{p}}{\gamma_{fast}\,V_{\rm fast}}$ up to $\sim 40$ for cases (A), (C), (D), while the 
outer jet has $M_{\rm fast}\sim 14.0$ in cases (A), (B2), (C), (D) and $M_{\rm fast}\sim 35$ in case (B1). 
However, both components are subsonically rotating, and during a complete rotation of the
inner jet, the fast magnetosonic or sound wave will propagate about $200$ times from the edge of the inner jet to the axis and back. 
Such overall configuration can easily develop non-axisymmetric instabilities, with growth times of order of the radial sound-crossing time~\citep{Hardee04}.

The computational domain of this simulation is a 2D box of size $-0.3 {\rm pc} <x<0.3 {\rm pc}$ and $-0.3 {\rm pc} <y<0.3 {\rm pc}$.
The simulation is performed in Cartesian coordinates using an HLLC flux formula \citep{Mignone&Bodo05}.  HLLC is a two state extension of the Harten, Lax and van Leer flux formulation (HLL), which includes a proper representation for the contact wave. We use a piecewise parabolic method (PPM) limiter \citep{Mignoneetal05}.  The combination of PPM reconstruction (third order accurate) and HLLC flux computations is extremely robust and handles both sharp discontinuities and turbulence development accurately. The lateral boundaries assume open boundary conditions, with a clipping of any inwardly directed momentum as soon as turbulent flow features start crossing the boundaries.
The simulation is run till time $t=50$, which due to our normalization translates to 163.2 year. The
corresponding distance of jet propagation of the jet beam during this time is about $50 {\rm pc}$.
The simulation is done using the AMRVAC code \citep{Melianietal07a,vanderHolstetla08} with Synge-type equation of state.
We take a base resolution of $120\times 120$, allow for 5 grid levels, reaching an effective resolution of $1920^2$. Shorter timescale runs at even higher resolutions were done to confirm that the dominant initial large scale structure development is 
adequately resolved, although more fine scale features inevitably turn up.
The simulations are performed typically using $120$ processors for about 2 days per case. 
We add some  white noise both at the initial time and at time $t=1$ when various waves have already developed  (this latter addition may not be essential to the evolution).
The fact that we use cartesian coordinates could give preference to instabilities with mode number 
proportional to $m = 4$ character (the case studied in \cite{Meliani&Keppen07b} was therefore confirmed
separately in cylindrical coordinates,  with the overall mode number dominance of $m=4$ recovered). However, in all cases we study here, the instabilities dominating 
the dynamics and evolution of the jet have a clear mixture of many mode numbers. 

\section{Results}\label{Simulations}

In the actual simulations performed in this work, the magnetic field in the jet is taken purely poloidal.
As explained before, our five simulations differ in their magnetic configuration: (A) is hydrodynamic; (B1) and (B2) has only the inner jet magnetized; (C) has only the outer jet magnetized; while in (D), both outflows are magnetized. This represents the main difference between the five cases, and while all
cases will develop fairly complex nonlinear evolutions governed by multiple, interacting instabilities, our main aim is to determine which
configuration can result in a clearly sustained two-component jet flow over a sufficiently long distance (time). 
It will turn out that cases (A), (C), and (D) all result in 
deceleration due to mixing between the two components, leading to decollimation of the jet, while only 
cases (B1) and (B2) convincingly maintain their two-component character. 
 For later reference, important models parameters are listed in Table~\ref{table:1}.
We now continue to discuss the complex nonlinear evolution of the five cases in some detail.

\begin{table*}
\begin{minipage}[t]{2\columnwidth}
\caption{The most relevant characteristics and parameters for all models investigated. In addition to these tabulated values,
the number density of the inner jet is fixed at $6.92\times 10^{-2} {\rm cm}^{-3}$, while $V_{\rm z}=0.99939 c$ and $v_{\rm \varphi}=0.01c$. Constant values for the outer jet are number density $119.94 {\rm cm}^{-3}$, with $V_{\rm z}= 0.9428 c$ and $v_{\rm \varphi}= 10^{-3} c$. The external medium is always static and unmagnetized, and has number density $10^{-2}{\rm cm}^{-3}$.}
\label{table:1}      
\centering                                      
\renewcommand{\footnoterule}{}
\begin{tabular}{c c c c c}
\hline
 &\multicolumn{2}{c}{inner jet}& \multicolumn{1}{c}{outer jet}&\multicolumn{1}{c}{effective inertia ratio}\\
\hline
case& $p_0$ & $B_{\rm z,in} (0.1375 {\rm Gauss})$& $B_{\rm z,out}$& $\gamma^2\rho h+B_z^2$ ratio out/in\\
\hline
\hline
A& 2.3 & 0.0&  0.0  & 0.15 \\
B1& 0.1 & 0.789& 0.0& 3.2\\
B2& $10^{-4}$ & 2.28 & 0.0& 18.3\\
C& 2.3  & 0.0& 2.323& 0.15\\
D&  2 & 0.789&  2.323& 0.17\\
\hline
\end{tabular}
\end{minipage}
\end{table*}

\begin{figure}
\begin{center}
\FIG{
{\resizebox{0.95\columnwidth}{0.78\columnwidth}{\includegraphics{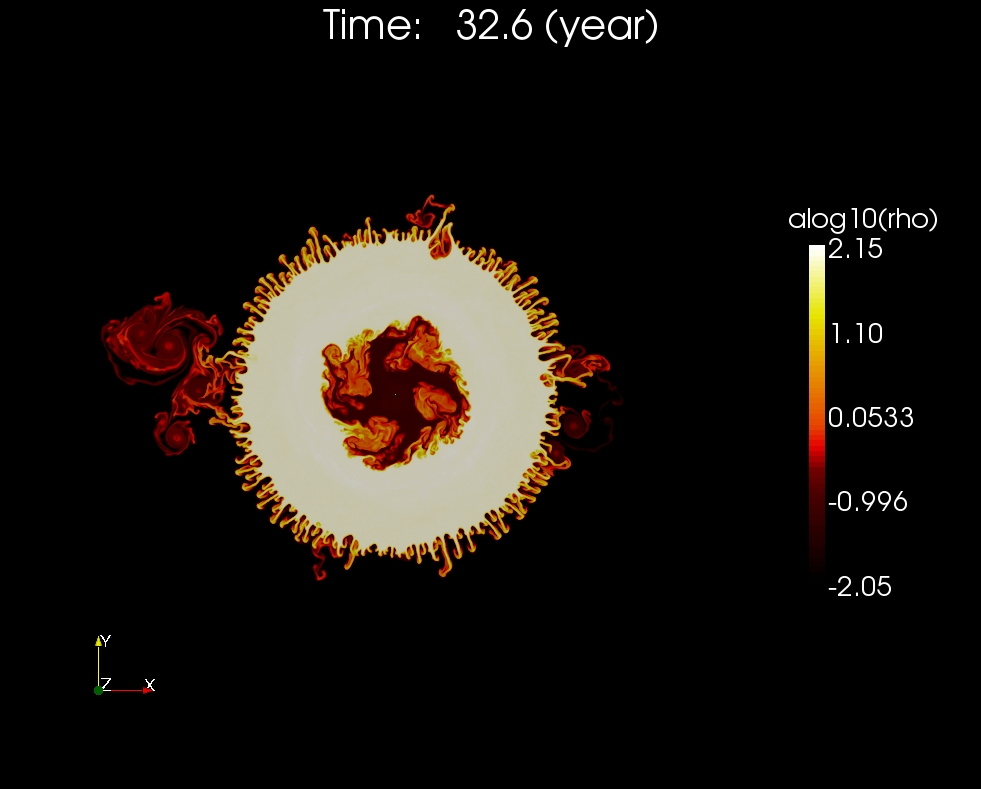}}}
{\resizebox{0.95\columnwidth}{0.78\columnwidth}{\includegraphics{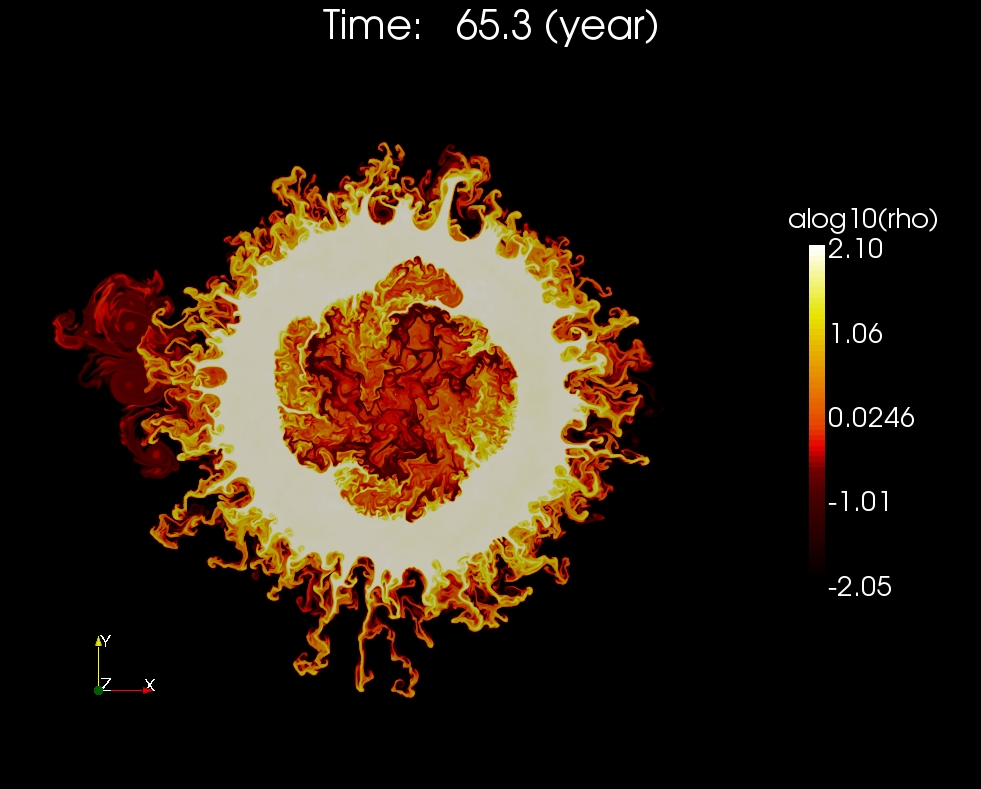}}}
{\resizebox{0.95\columnwidth}{0.78\columnwidth}{\includegraphics{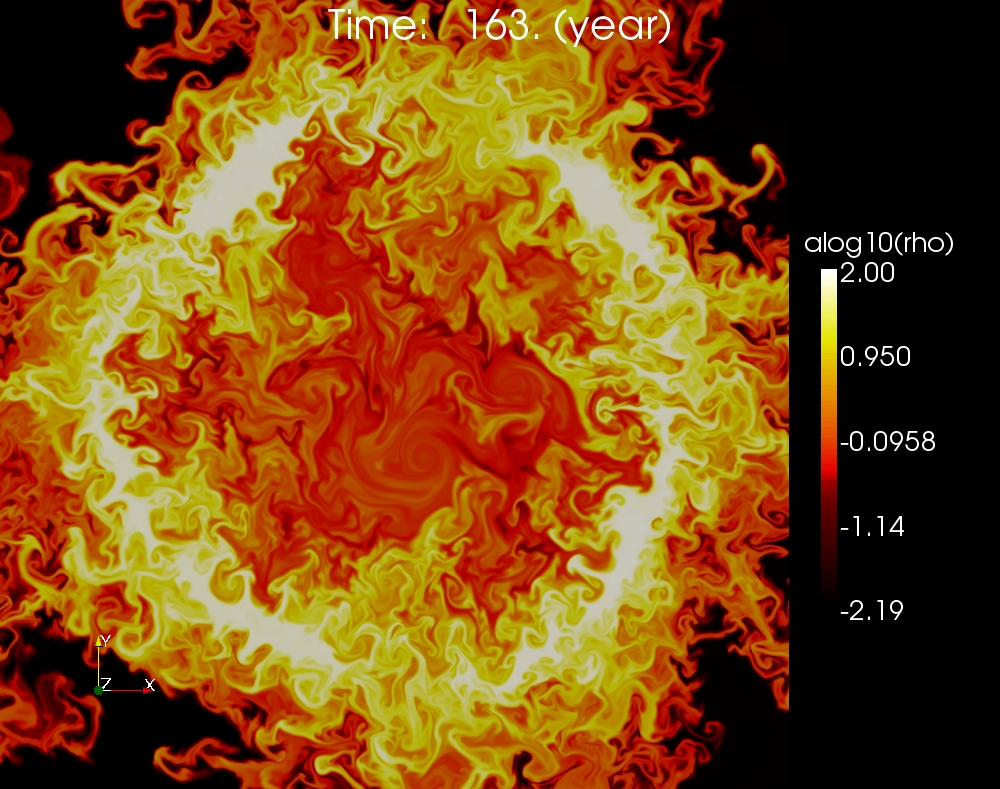}}}
}
\caption{Case (A), the purely hydrodynamical jet, showing logarithm of proper density at times (top) $t=32.6 {\rm year}$: developement of relativistically enhanced Rayleigh-Taylor-type instability propagating inward in the inner jet, (center) $t=65.3{\rm year}$: merging of the three Rayleigh-Taylor fingers at the jet axis and (down) $t=163 {\rm year}$: decollimation of the jet} (one full rotation of the inner jet corresponds to $t=65.3 {\rm year}$).\label{figA}
\end{center}
\end{figure}
\subsection{Case (A)}\label{caseA}

According to the Rayleigh criteria mentioned, the interface between the two components is always 
unstable. Also, the effective inertia of the inner jet is higher than in the outer jet. 
In fact, at the interface $R=R_{{\rm in}}$, we have initially $\left(\gamma^2\rho\,h\right)_{\rm out}\approx 0.07\left(\gamma^2\rho\,h\right)_{\rm in}$. In Table~\ref{table:1}, a $t=0$ ratio is given using a value midway the outer jet divided by the axial value. The overall effective inertia contrast is very different from the case we investigated in \cite{Meliani&Keppen07b} ,
where the ratio between (mean value of) effective inertia of the inner component jet to the effective inertia of 
the outer component jet was such that $\left(\gamma^2\rho\,h\right)_{\rm out} \gg \left(\gamma^2\rho\,h\right)_{\rm in}$. 

Initially, a linear surface mode develops at 
this interface $R=R_{{\rm in}}$. As a direct consequence, a small, radially extended shear region with low effective inertia $\gamma^2\rho\,h$ forms at this location. 
At both bounding interfaces of this small shear layer, one with the outer and one with the inner jet, small wavelength instabilities develop. 
Meanwhile, acoustic waves from the inner interface propagate inwards, reflect on the jet axis, and then give rise to 
a Kelvin-Helmholtz-like body 
 mode with dominant azimuthal mode number $m\sim 8$ in the inner jet. 
Meanwhile, the surface instability at the interface between inner jet and shear shell reaches a clearly non-linear phase. 
The overall interaction then gives rise to the growth of a relativistically enhanced Rayleigh-Taylor-type instability, having a body mode 
character propagating into the inner jet (Fig.~\ref{figA}, top). This instability regime shows about four arms at first, that collapse to three arms which progressively propagate inwards. As will be detailed
later on, this instability can be explained from the fact that in this case (A), fluid elements in the shear layer have a lower effective 
inertia $\gamma^2\rho\,h$ and higher pressure than pre-existing material in the inner jet.

We could qualitatively describe this relativistic Rayleigh-Taylor instability as follows. 
A fluid element in the shear layer is at first typically rotating with a speed of order of $V_{\varphi}\sim 0.01$, the initial rotation speed at the inbound 
interface. The lower inertia of the shear shell relative to the inner jet makes that transfer of angular momentum from the inner wind/jet region to the shear layer is
rather efficient, causing a fast rotational speedup of the shear layer matter. 
At the same time, the centrifugal force acting on a fluid element moving inwards varies with $1/R$. 
Because the fluid inertia in the shear region ends up at about a twelve times lower value than the inertia in the inner jet, 
the centrifugal force acting on this fluid element is much weaker than in the inner wind. 
At the same time, the pressure in the shear layer fluid element is higher than in the inner wind. 
Therefore, the element will expand inwards until the pressure gradient gets balanced by the centrifugal force.  
In this particular simulation, three prominent fingers form and start propagating inwards.
However, as the initial pressure profile decreases towards the jet axis, the centrifugal force in the inner jet decreases inward. 
With such a configuration, we find that in the phase where the intruding, rotating fingers of the Rayleigh-Taylor mode propagate inward, 
they will nearly reach the jet axis.  
Very close to the axis, the centrifugal force acting on them gets compensated by 
the pressure force. In the phase that follows, the rotating fingers get deflected sideways in the direction of rotation. 
After less than half a full rotation of the inner jet (at $t\sim 32.6 {\rm year}$, see Fig.~\ref{figA}, top), 
this relativistic Rayleigh-Taylor mode then dominates the body mode 
instability in the inner jet. The Rayleigh-Taylor fingers propagating inwards also compress the inner jet. Since angular momentum is conserved,
the centrifugal force increases and the inner jet fluid moves outwards towards the shear shell. This means that we get enhanced angular momentum 
transfer from the inner jet to the shear layer shell, which accelerates the rotational speed of the shear shell.
We find that after less than one total rotation of the inner jet (at $t\sim 65.3 {\rm year}$, see Fig.~\ref{figA}, center), the three Rayleigh-Taylor fingers nearly merge, making the inner jet appearance
totally dominated by the (growing) shear shell. 

An analysis of this relativistic Rayleigh-Taylor-type instability is described seperately in Section~\ref{lninst}.
The low effective inertia of fluid in the shear region makes the interface between the shear flow and the inner jet unstable to this instability, where centrifugal forces act as an effective (radially outwards pointing) gravity. Once initiated, these grow faster than the previously formed 
Kelvin-Helmholtz body mode. A number of `arms' develop from the shear region, having lower angular momentum than the fluid from the inner region, these propagate
inwards, while inner jet fluid having higher angular momentum propagates outwards. The end result is transfer of energy and angular momentum from 
the inner jet to the growing shear region. As a consequence of this complex mixing of jet and shear layer matter,
we find significant deceleration of the inner jet, while the density and the radius of the inner jet plus 
shear region increases (this is quantified later in Fig.~\ref{Inradius_evolution}).

The external boundary of the growing shear region at $R\approx R_{\rm in}$ which borders the outer cold jet is also Kelvin-Helmholtz unstable. For the case at hand, in a first phase, a larger scale 
Kelvin-Helmholtz type mode develops with 8 undulations initially, eventually converging to form about 5 large 
scale structures (see Fig.~\ref{figA}, center). This is different
from the case studied before
in~\cite{Meliani&Keppen07b}, where the most prominent feature in the non-linear evolution was the development of
a Kelvin-Helmholtz type mode with $4$ arms. This mode
transfers angular momentum from the shear region to the outer jet and decollimates the jet. At smaller spatial scale, but also at the interface 
between the shear region and the outer jet, other instabilities develop as well. These are again more centrifugally driven, as angular momentum and pressure increase in 
the shear region. Thus also at this interface, the centrifugal force becomes locally higher than in the outer outflow. Small bubble-like protrusions from the shear shell 
enter the outer outflow. However, since the pressure of the outer jet slowly increases outward, the growth of this smaller scale instability is stopped when the 
centrifugal force of the protruding bubbles gets balanced by the pressure of the outer jet. Therefore, this instability remains weak and the large-scale 
dynamics of the interaction between the shear layer and the outer jet remains dominated by the Kelvin-Helmholtz instability. 
The small scale instabilities do influence the dynamics, by increasing the efficiency of the angular momentum transfer from the Kelvin-Helmholtz structures to 
the outer jet. Finally, also the outer interface  at $R\approx R_{\rm out}$ between outer jet and the hot, dilute medium is 
Rayleigh-Taylor unstable, and is dominated by small wavelength perturbations.

As the overall outcome of all the interacting instabilities, the two-component jet decollimates, such that
the jet radius increases to $R_{\rm out}\sim 0.28 {\rm pc}$ (see Fig.~\ref{Radius_evolution}). The inner 
component jet also spreads and its radius reaches $R\sim 0.18 {\rm pc}$ (see Fig.~\ref{Inradius_evolution}). 
Also during the simulated $163$ years in physical time (Fig.~\ref{figA}, down), the rotation speed of both
 components decreases 
and the interaction between the shear region and the outer component is dominated by Kelvin-Helmholtz induced behaviour, together with the relativistically enhanced 
Rayleigh-Taylor instability of the inner rotating fluid. 
The main result of this interaction between the two components is the dramatic deceleration of the inner jet. 
In fact, the relativistic Rayleigh-Taylor-type instability leads to a deceleration of the inner component 
where the Lorentz factor of the inner component drops to $\gamma\sim 8$ (quantified later for all models in Fig.~\ref{Lorentzfactor_evolution}).

\subsection{Cases (B1) and (B2)}\label{caseB}

\begin{figure}
\begin{center}
\FIG{
{\resizebox{0.95\columnwidth}{0.78\columnwidth}{\includegraphics{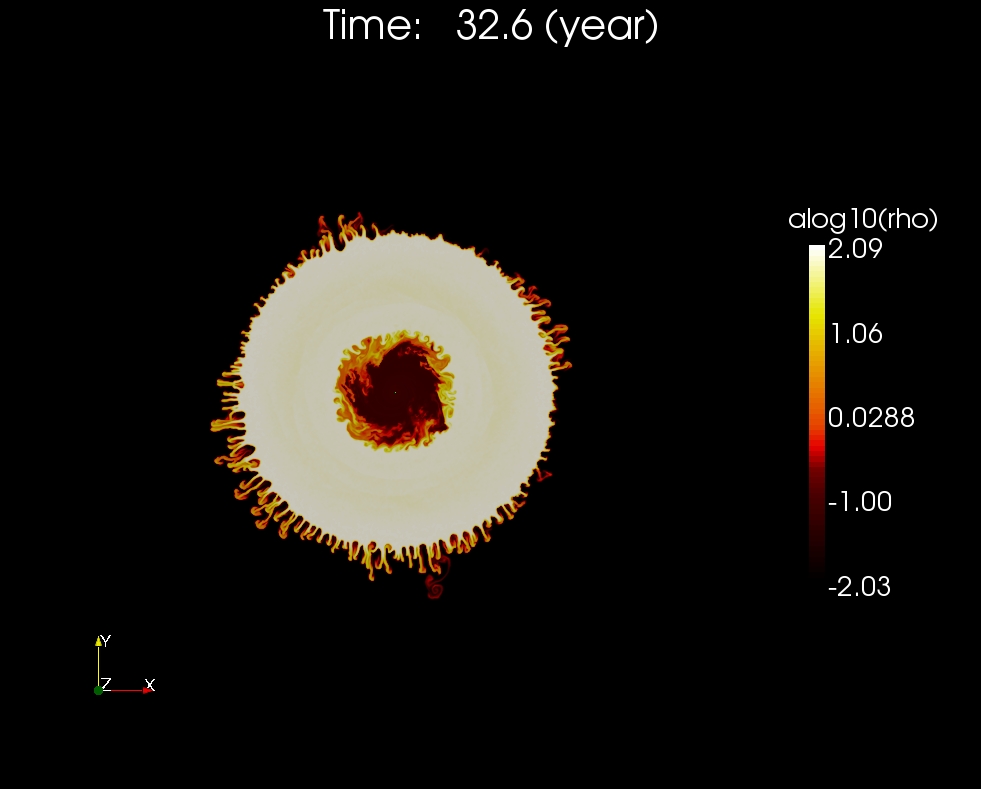}}}
{\resizebox{0.95\columnwidth}{0.78\columnwidth}{\includegraphics{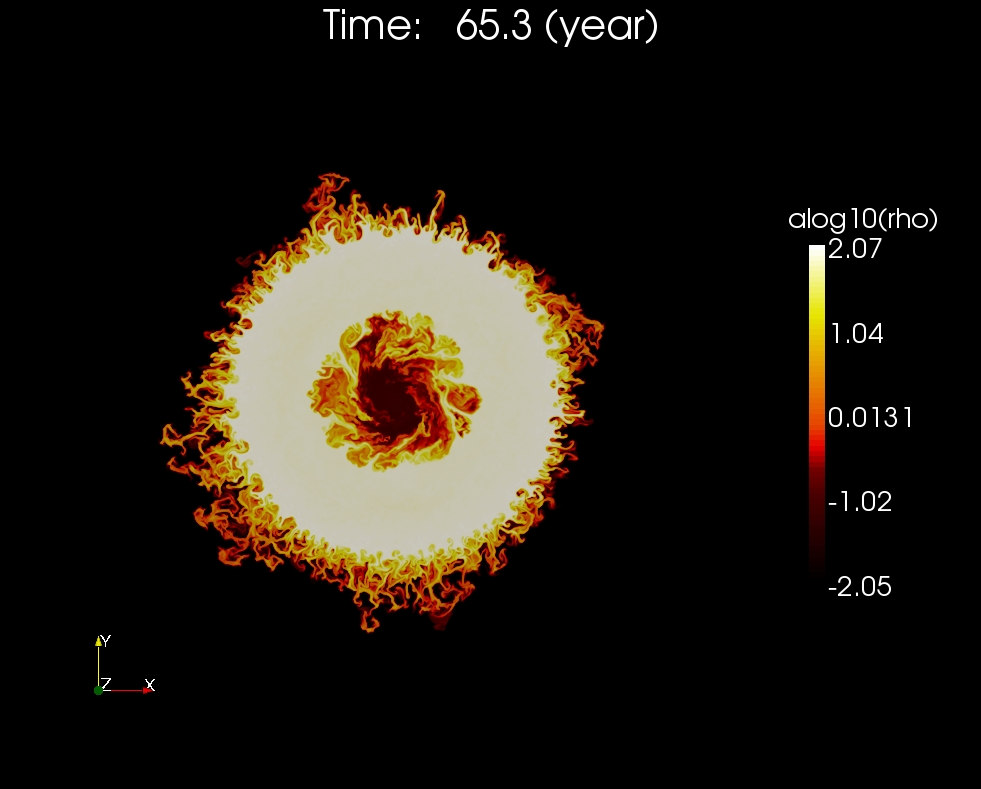}}}
{\resizebox{0.95\columnwidth}{0.78\columnwidth}{\includegraphics{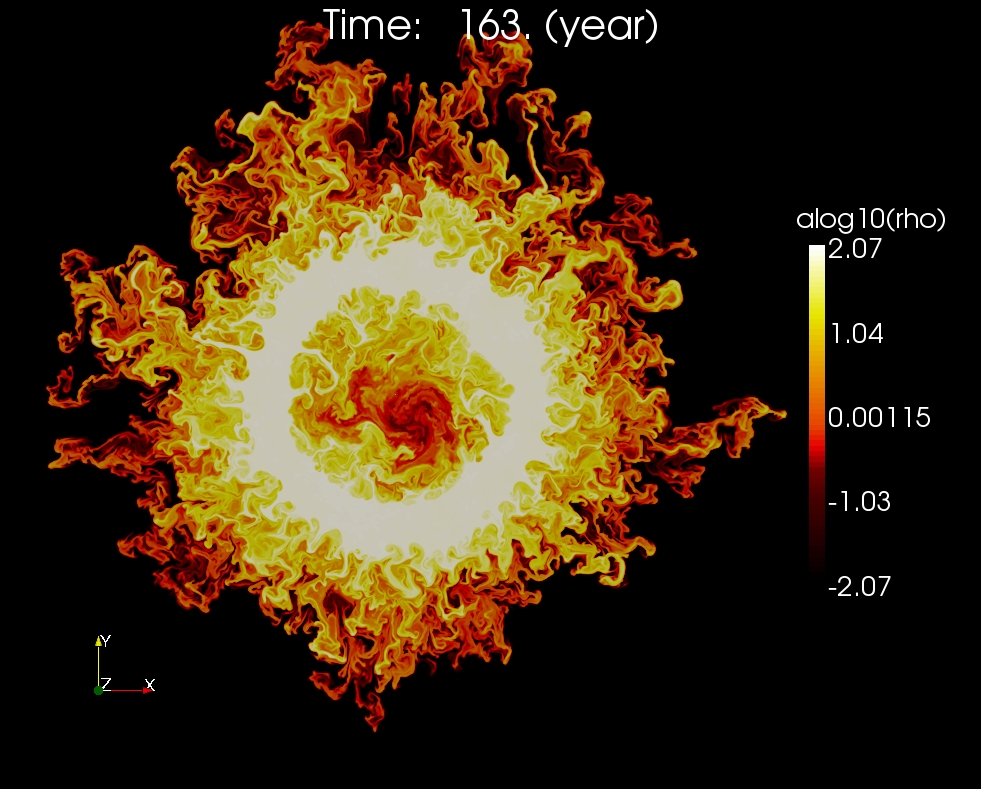}}}
}
\caption{Case (B1), only the inner jet is magnetized. Logarithm of density at (top) $t=32.6 {\rm year}$: developement of Kelvin-Helmholtz body mode instability at the  interface between the inner and outer jet, (center) $t=65.3 {\rm year}$: inward growing of the shear shell in the inner jet and (down) $t=163 {\rm year}$: developement of Rayleigh-Taylor at the outer jet interface, with one rotation of the inner jet completed at $t=65.3 {\rm year}$.}\label{figB}
\end{center}
\end{figure}

\begin{figure}
\begin{center}
\FIG{
{\resizebox{0.95\columnwidth}{0.78\columnwidth}{\includegraphics{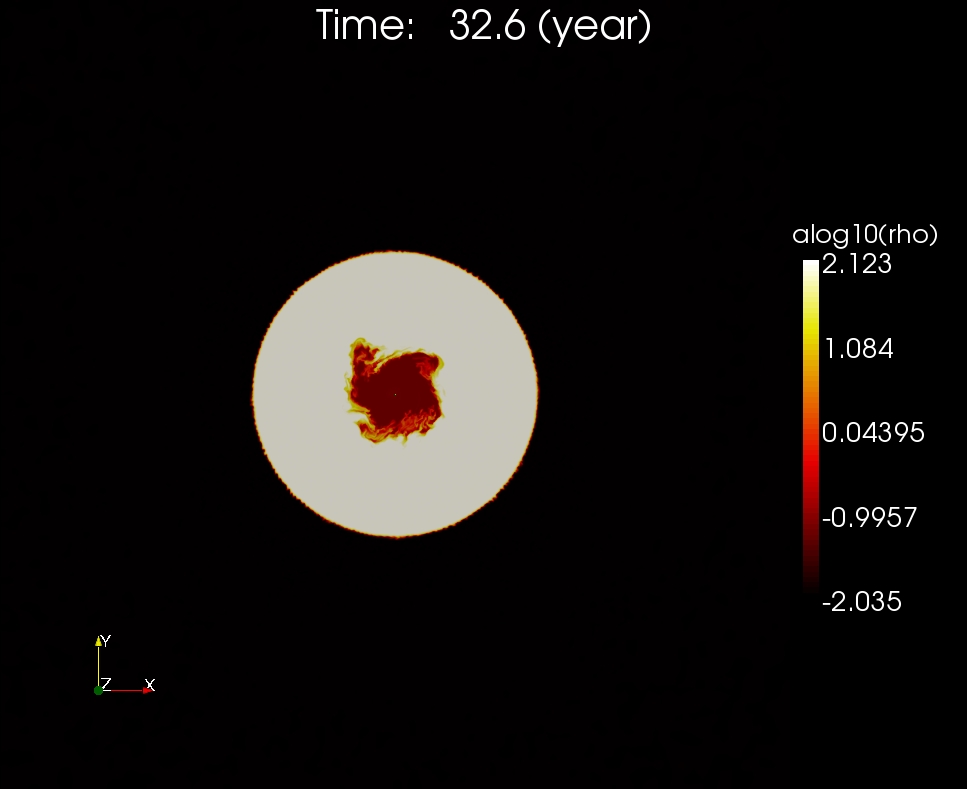}}}
{\resizebox{0.95\columnwidth}{0.78\columnwidth}{\includegraphics{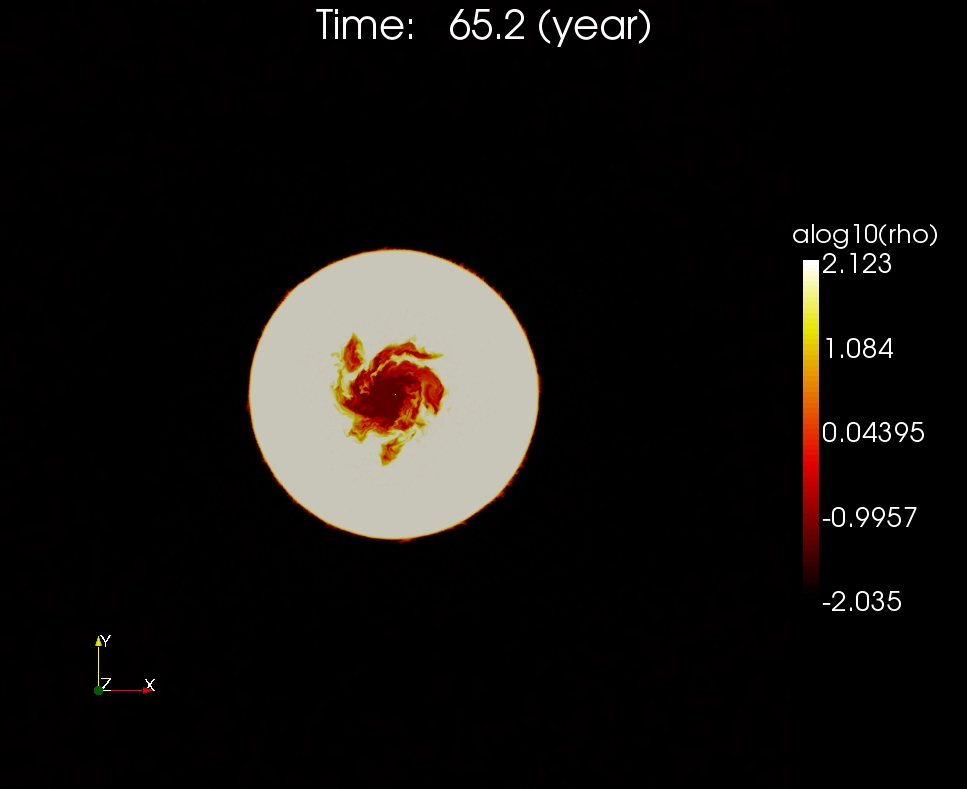}}}
{\resizebox{0.95\columnwidth}{0.78\columnwidth}{\includegraphics{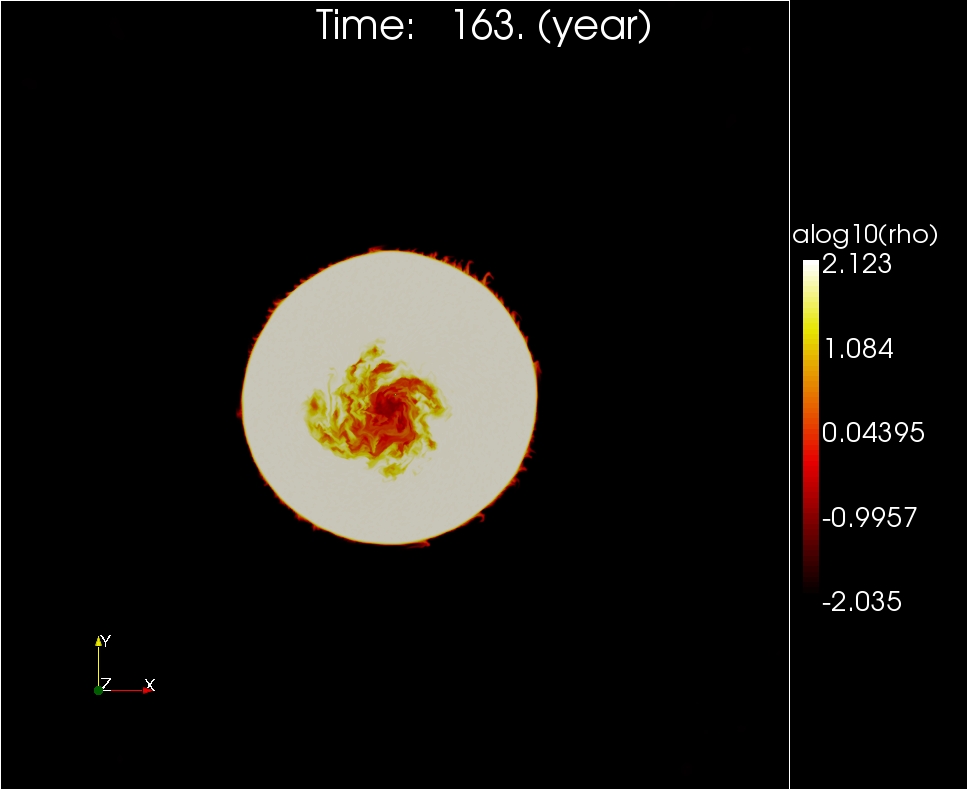}}}
}
\caption{Case (B2), only the inner jet is magnetized. Logarithm of density at (top) $t=32.6 {\rm year}$: developement of Kelvin-Helmholtz body mode instability at the  interface between the inner and outer jet, (center) $t=65.3 {\rm year}$ and (down) $t=163 {\rm year}$: the inner jet component is deplaced  from its on-axis position due to significant non-axisymmetric mode development, with one rotation of the inner jet completed at $t=65.3 {\rm year}$.}\label{figB2}
\end{center}
\end{figure}

We now describe the two-component jet evolution with a magnetized inner component. 
The effective inertia in the inner jet is lower than in the outer jet with initial contrasts $\left(\gamma^2\rho\,h\right)_{\rm out}\approx 3.2 \left(\gamma^2\rho\,h+B_{\rm z}^2\right)_{\rm in}$ in case (B1) and 
$\left(\gamma^2\rho\,h\right)_{\rm out}\approx 18 \left(\gamma^2\rho\,h+B_{\rm z}^2\right)_{\rm in}$ 
in the case (B2), due to the initial lower pressure at the jet axis and the initial different  distribution over thermal and magnetic energy in them. 
These cases are then such that the inner interface is now stable against the dominating Rayleigh-Taylor type instability described above and explained
analytically in Section \ref{lninst}. Still, the interface  at $R\approx R_{\rm in}$ between the inner and outer jet is subject to Kelvin-Helmholtz instability because of the differential
rotation. This modifies the shape of the interface surface, leading to
non-axisymmetric reflection of magnetohydrodynamic waves propagating through the inner jet. These waves in turn disturb locally the initial equilibrium between 
total pressure gradient and centrifugal force.

During the evolution, some spikes develop at the interface, where the outer jet locally interchanges with the inner jet. These spikes are accelerated in the
toroidal direction by the faster rotating inner jet. The centrifugal force acting on these slows down their inward expansion, and they then mainly propagate in the 
toroidal direction. 
Their interaction with the inner jet material induces Kelvin-Helmholtz body mode instability, with a spiral 
pattern forming, this time having 3 arms in the two cases (see Fig.~\ref{figB}, top and Fig.~\ref{figB2}, top). 
They extract some angular momentum from the outer regions of the inner jet, slowing its rotation. Again, a shear shell then forms in this region, continuously extracting 
angular momentum from the inner jet/wind. The spikes then slowly expand to the jet axis. The inner wind gets compressed by the shear layer. 
However, these mainly Kelvin-Helmholtz type instabilities turn out to have lower efficiency in extracting angular momentum than the relativistic Rayleigh-Taylor 
instability encountered in case~(A). The compression of the inner jet by the shear region is followed by a 
modest expansion. In the case (B2), the extraction 
of the angular momentum is weaker than in the case (B1), because in case (B2) the inner jet has lower inertia.
Meanwhile, the angular momentum extracted from the inner jet leads to an outwards extension of the shear shell  that formed at $R_{\rm in}$. At the interface between this growing shear shell and the outer jet, a Kelvin-Helmholtz instability surface mode with azimuthal mode number $m\sim 3$ character then develops. Also at this interface, smaller scale
Rayleigh-Taylor instability proceeds. The shear shell has an effective inertia which is lower than the outer jet inertia. 
As locally centrifugal forces dominate, low inertia `bubbles' radially extend toward the outer jet.  
The growth rate of these bubbles is rather slow, since the pressure of the outer jet increases mildly outwards and the slowly rotating outer jet breaks their rotation.

At the outer surface  $R_{\rm out}$ of the outer jet in the case (B1) (see Fig.~\ref{figB}), we also witness 
Rayleigh-Taylor instability, as in case (A). On the other hand, case (B2) shows little evidence for
such instability at the outer jet surface (see Fig.~\ref{figB2}), which is due to the different external pressure conditions there. 
Throughout the simulation, the jet in (B1) and (B2) remains collimated by the outer component, which 
compresses the inner component and shear region. The inner component jet radius in case (B1) at $163$ years 
remains lower than $0.07 {\rm pc}$ and in case (B2) even below about $0.04 {\rm pc}$ (see
Fig.~\ref{Inradius_evolution}). 
However, in case (B1) the Rayleigh-Taylor instability at the surface of the outer region forms an extended sheat,
which gives an increase to the apparant total jet radius $R_{\rm jet}=0.2 {\rm pc}$ 
(see Fig.~\ref{Radius_evolution}), whereas the jet in case (B2) does not show any sign of decollimation.

Both the inner jet and the shear region end up magnetized. The main difference between these cases and all others studied here, is that despite all small-scale instability development,
the inner jet decelerates little, dropping to an average Lorentz factor of about $20$ (see
Fig.~\ref{Lorentzfactor_evolution}). The two-component jets for case (B1) and (B2) remain clearly
separable in inner and outer jet components (see Figs.~\ref{figB}, down-\ref{figB2}, down), which is not the case for 
all other evolutions shown here. In both cases (B1) and (B2), we do find that the inner jet component is deplaced 
from its on-axis position due do significant non-axisymmetric mode development. 
The jet stratification converges to a structure with an inner fast, magnetized spine having a Lorentz factor of 
about $20$. The spine is surrounded by a shear shell being $100$ denser than the spine and with lower Lorentz factor and low magnetisation.
The difference between the two cases is that the high magnetic pressure and low thermal pressure in the inner jet for case (B2) increases its non-linear stability as compared to case (B1).

\subsection{Case (C)}\label{caseC}

\begin{figure}
\begin{center}
\FIG{
{\resizebox{0.95\columnwidth}{0.78\columnwidth}{\includegraphics{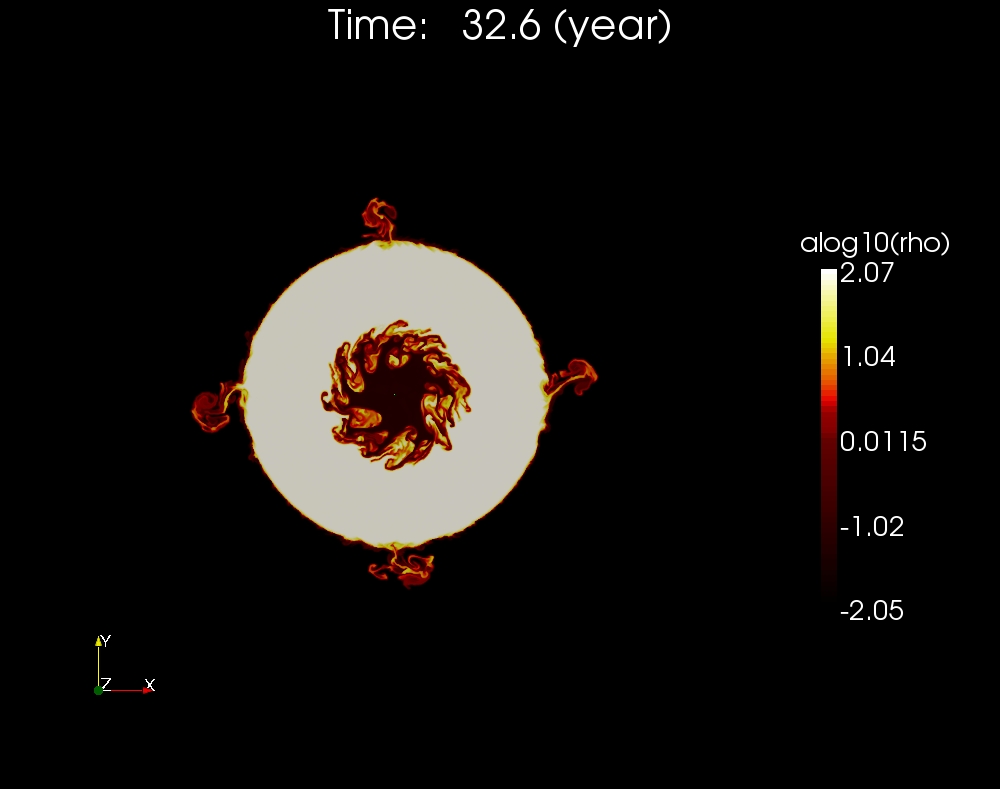}}}
{\resizebox{0.95\columnwidth}{0.78\columnwidth}{\includegraphics{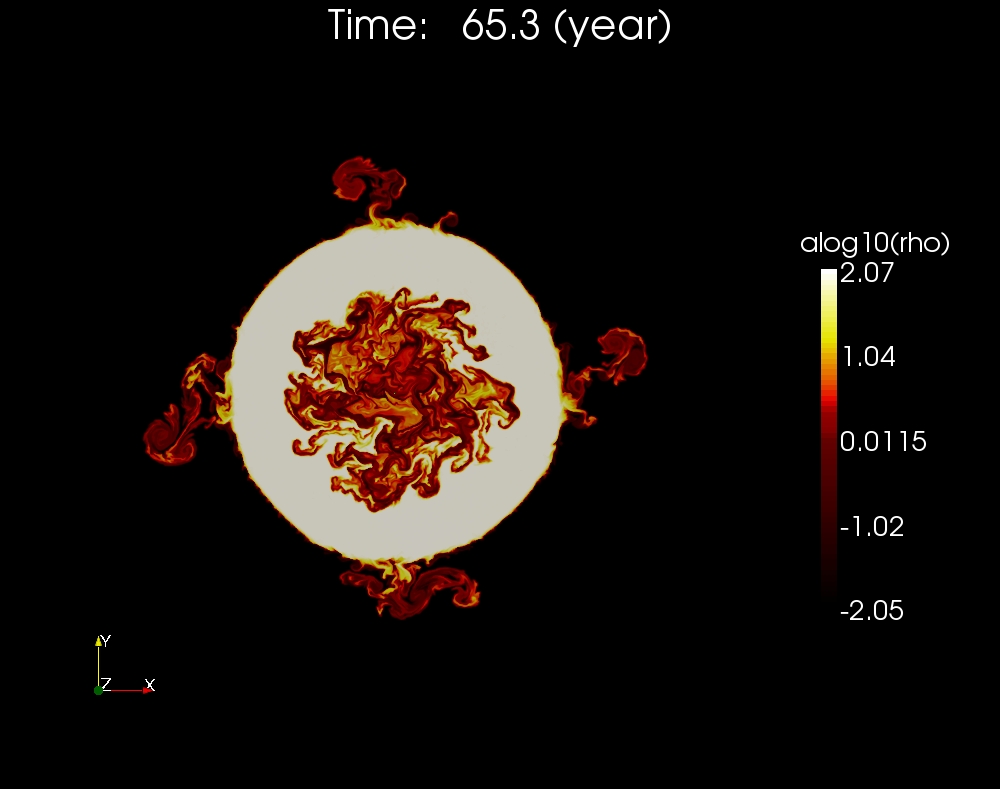}}}
{\resizebox{0.95\columnwidth}{0.78\columnwidth}{\includegraphics{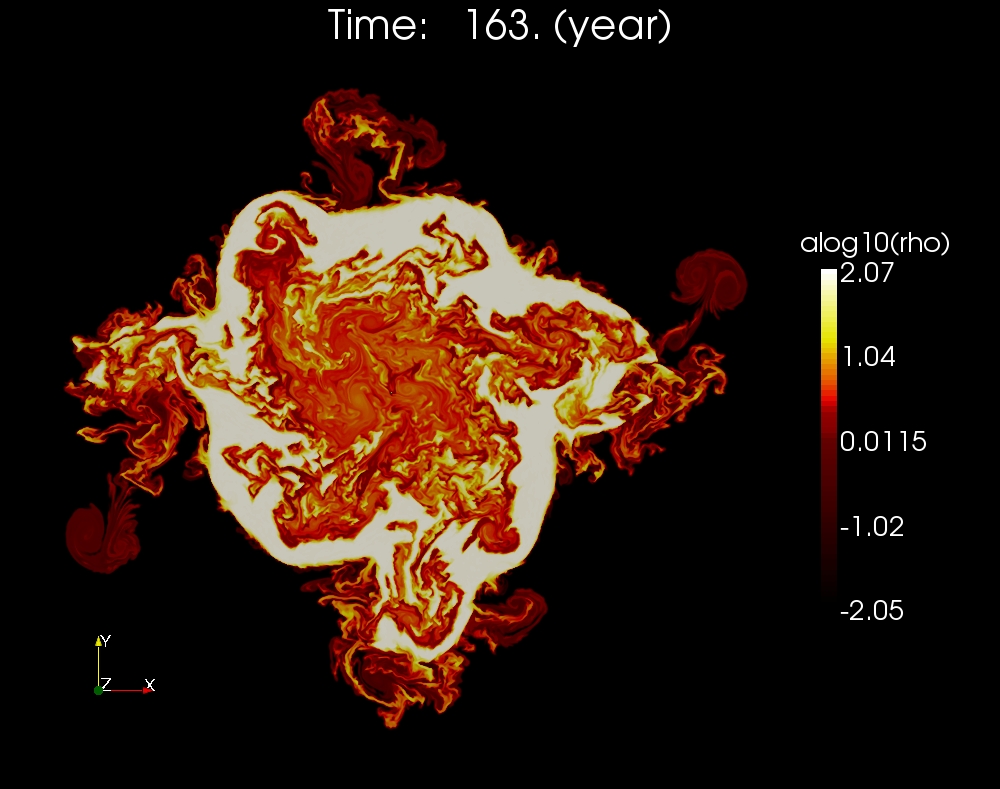}}}
}
\caption{Case (C), only the outer jet is magnetized. Logarithm of density at (top) $t=32.6 {\rm year}$: Rayleigh-Taylor instability develops at interface between the two component inducing rotating `bubbles' that protrude inward, (center) $t=65.3 {\rm year}$: `bubbles' merge at the jet axis and inner jet decelerates and becomes turbulent and (down) $t=163 {\rm year}$: The shear shell expand outward in the outer jet and decollimates the jet, with one full inner jet rotation at $t=65.3 {\rm year}$.}\label{figC}
\end{center}
\end{figure}

In this case, only the outer jet is magnetized with a purely poloidal magnetic field.
As mentioned before, the rotation of the jet is subsonic, such that acoustic waves which start to propagate from the inner-outer jet interface at different stages in the evolution propagate inwards and are reflected about $200$ times during one rotation period of the inner jet. Our simulations follow about 2.5 full inner jet rotations.
As the interface between the two components is unstable to various mode types as already encountered in the previous cases, its shape changes and the acoustic wave propagation becomes strongly non-axisymmetric. These waves 
play a clear role in disturbing the initial axisymmetry of the inner jet, and mitigate the development of non-axisymmetric 
body mode instabilities in the inner jet. 

In this case~(C), in a first stage, linear small-wavelength (compared to case (A)) Kelvin-Helmholtz instabilities develop at the interface $R\approx R_{\rm in}$ between the two components. These
reach a non-linear phase in less than a tenth of a full rotation. Spikes of the outer jet, which are this time magnetized, arise in the inner jet. 
The large ratio of the fluid inertia between the outer and inner jet makes these spikes once more unstable to the centrifugally mediated, relativistically enhanced 
Rayleigh-Taylor instability seen previously in case (A). This gives rise to the formation of rotating `bubbles' that protrude inward. In the case at hand, about ten such 
bubbles form (see Fig.~\ref{figC}, top), having smaller size compared to those seen to emerge in case (A). This difference results due to the difference in the Kelvin-Helmholtz surface mode development
mentioned earlier. The initial smaller size of the protruding bubbles makes their growth rate slower than in the purely hydrodynamical case (A), 
until they reach the merging phase where in this case about three larger size bubbles form and converge to the jet axis. Meanwhile, these bubbles compress 
and push fluid elements from the inner jet to a shear layer (see Fig.~\ref{figC}center). Thus once again, the larger-scale mode structures act to transfer angular momentum from the inner jet to 
the shear region.

Since the bubbles which move radially inwards form at the shear shell, 
their ratio of magnetic pressure to thermal pressure $B_{\rm z}^2\left(1-V_{\varphi}^2\right)/(2\,p)\sim 0.1$ 
is lower than in the outer jet.
In fact, the shear region is made up of fluid from the inner and the outer jet, and its thermal pressure dominates total pressure.
The external surface of the shear shell is Kelvin-Helmholtz unstable and about 6 arms form (see Fig.~\ref{figC}, center). From each one, a centrifugally driven spike propagates outward. 
In fact, angular momentum extracted from the inner jet in the end gets transferred to the external part of the shear region. Then, locally the centrifugal force at the
interface between the shear shell and the outer jet acts to develop spikes which decollimate the 
jet (see Fig.~\ref{figC}, down). The total jet radius increases from $0.1 {\rm pc}$ at the initial stage to $0.189 {\rm pc}$ (Fig.~\ref{Radius_evolution}), after the jet has propagated for a distance of $50 {\rm pc}$. 
In the same time, the inner jet spreads as well, decelerates and its Lorentz factor drops to about $8$.
At the endstate, the jet is constituted of an inner turbulent component with a radius of $0.15 {\rm pc}$ (see
Fig.~\ref{Inradius_evolution}) and a small outer component. 
The final rotation speed profile increases from the axis 
to $0.006$ at $0.04 {\rm pc}$, further decreasing outwards. 
\begin{figure}
\begin{center}
\FIG{
{\resizebox{0.95\columnwidth}{0.78\columnwidth}{\includegraphics{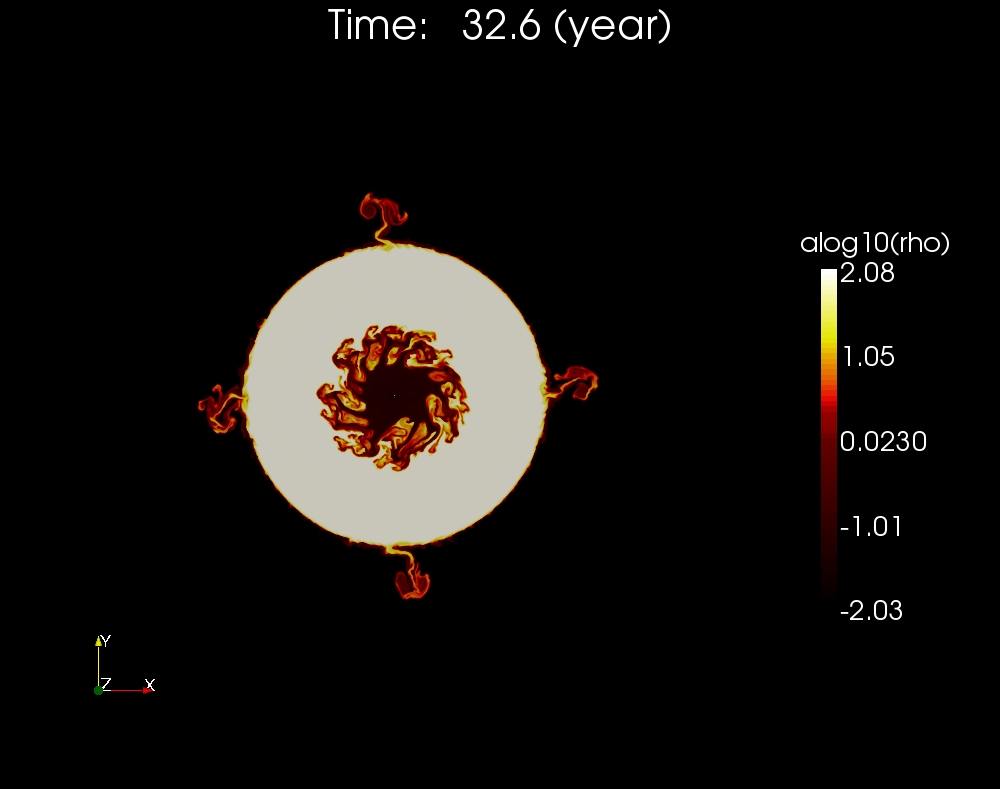}}}
{\resizebox{0.95\columnwidth}{0.78\columnwidth}{\includegraphics{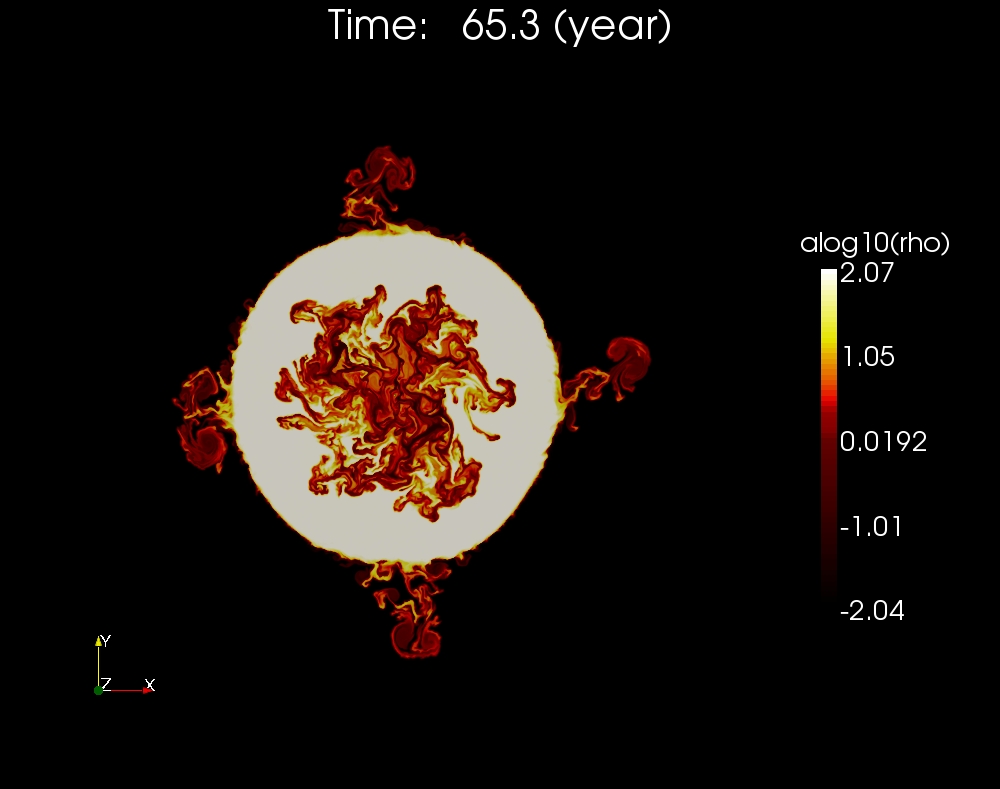}}}
{\resizebox{0.95\columnwidth}{0.78\columnwidth}{\includegraphics{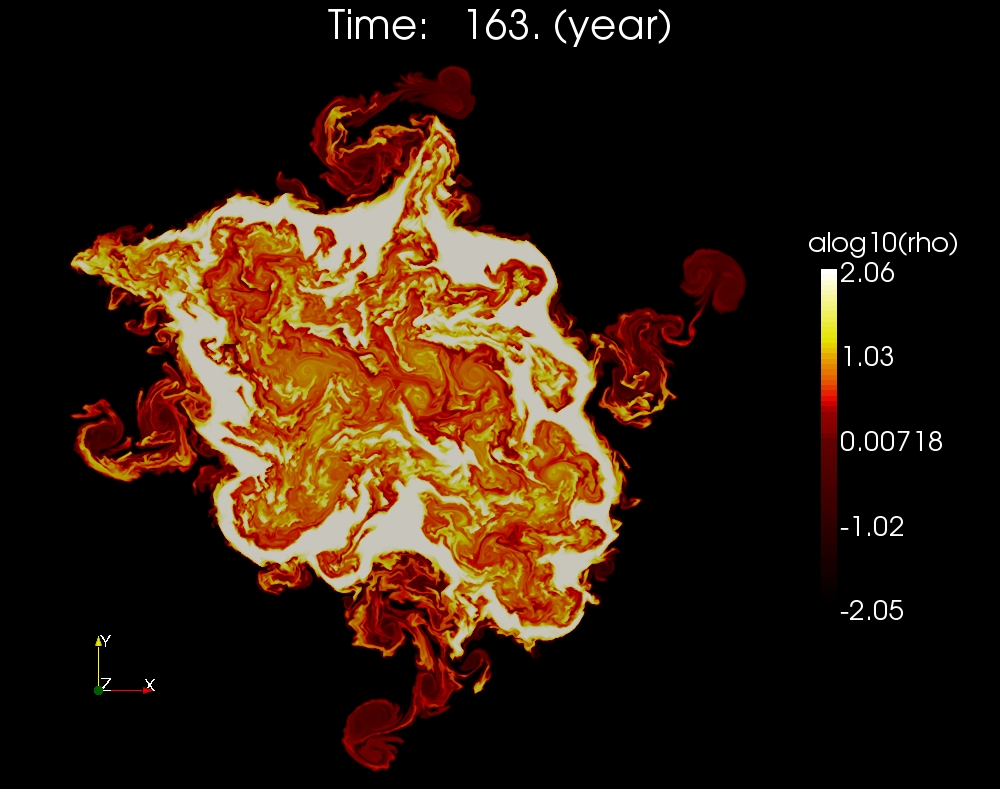}}}
}
\caption{Case (D), both jet components magnetized. Logarithm of density at (top) $t=32.6 {\rm year}$: 13 bubbles arise, 
moving inwards into the inner jet, (center) $t=65.3 {\rm year}$: Rayleigh-Taylor bubbles emerge at the jet axis and the inner jet becomes turbulent and (down) $t=163 {\rm year}$: The shear region grows and changes the full jet shape 
to elliptic, with one full rotation of the inner jet corresponding to $t=65.3 {\rm year}$.}\label{figD}
\end{center}
\end{figure} 
\subsection{Case (D)}\label{caseD}

\begin{figure}[h]
\begin{center}
\vspace{-1cm}
\FIG{
{\resizebox{0.95\columnwidth}{6cm}{\includegraphics{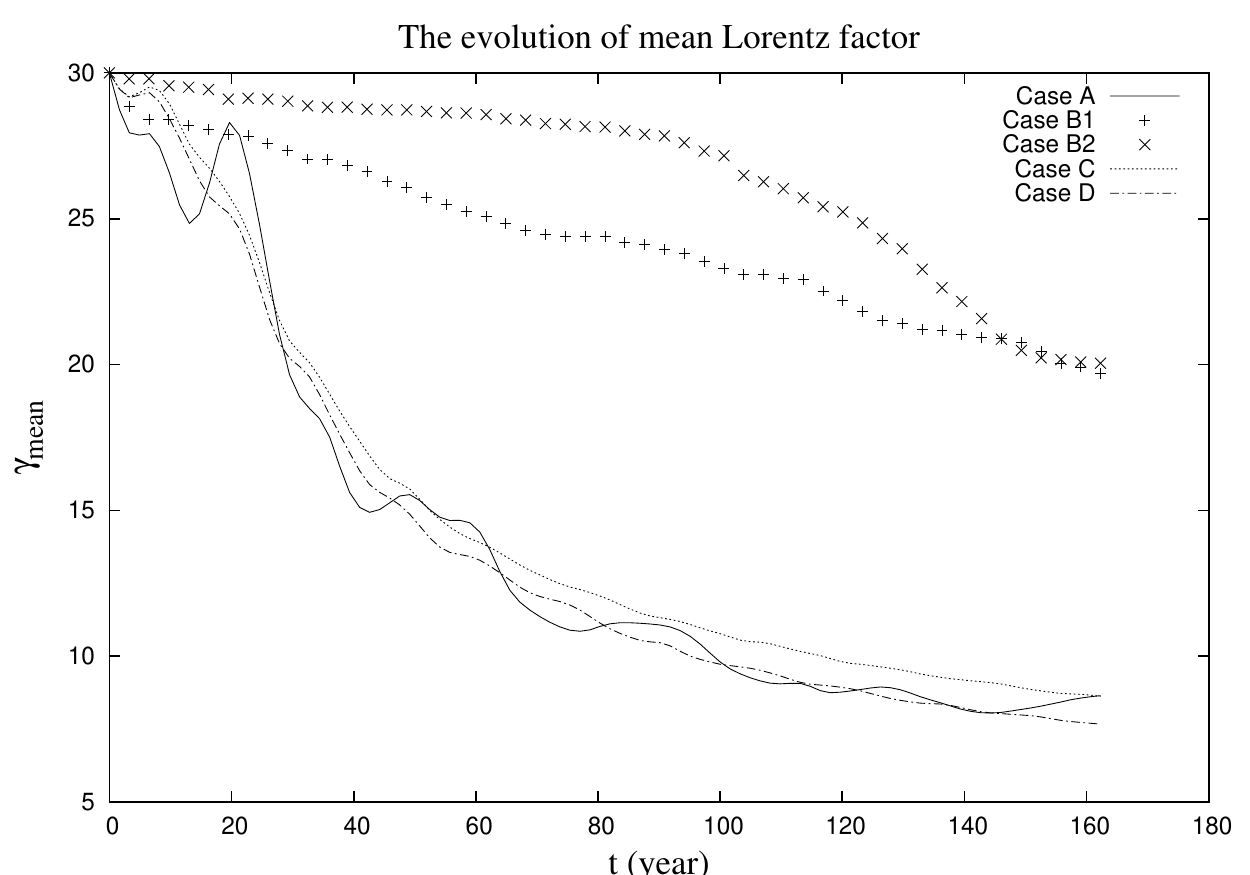}}}
}
\caption{Time evolution of the mean Lorentz factor of the inner jet in the five cases. 
For the cases (A), (C), and (D) we distinguish between the inner and outer jet according to the polytropic 
index and Lorentz factor. 
In the cases (B1) and (B2), we use the same condition for the Lorentz factor to distinguish between the inner 
and outer jet, as well as
the extra condition that the inner jet is magnetized.}\label{Lorentzfactor_evolution}
\vspace{-0.1cm}
\end{center}
\end{figure}

In this last case, both inner and outer jet are magnetized. 
Initially, from the discontinuity between the two components, fast magnetosonic waves propagate both outwards into the outer jet, and inwards into the inner jet. 
The end result of the initial readjustment is that a small shear region with lower magnetization than both component develops at the interface  $R\approx R_{\rm in}$. Also this configuration
is Kelvin-Helmholtz unstable. At the surface from the shear region roughly 13 bubbles arise, 
moving inwards into the inner jet (see Fig.~\ref{figD}, top). They merge to form 3 large arms 
expanding towards the jet axis (see Fig.~\ref{figD}, center). Fluid from the inner jet is pushed outwards, thereby extracting angular momentum from the inner jet. Since the structures form at the shear region, 
the magnetic pressure inside the bubbles is lower than in the inner and outer component. However, the thermal pressure is increased in the shear region. 
During the evolution, dominant forces are magnetic pressure and thermal pressure that balance each other, while the contribution of the centrifugal force is low. 
However, the centrifugal force remains responsible for triggering instabilities. As in case (C), the shear region grows with eventual almost complete destruction of the 
inner jet component (see Fig.~\ref{figD}, center). In the last phase simulated, the dominating inner shear region becomes totally turbulent, and strong vortices form, changing the full jet shape 
to elliptic (Fig.~\ref{figD}, down). At this phase the average Lorentz factor of the inner jet has dropped down to
$\gamma_{\rm max}\sim 7$ (see Fig.~\ref{Lorentzfactor_evolution}).

\begin{figure}[h]
\begin{center}
\vspace{-1cm}
\FIG{
{\resizebox{0.95\columnwidth}{6cm}{\includegraphics{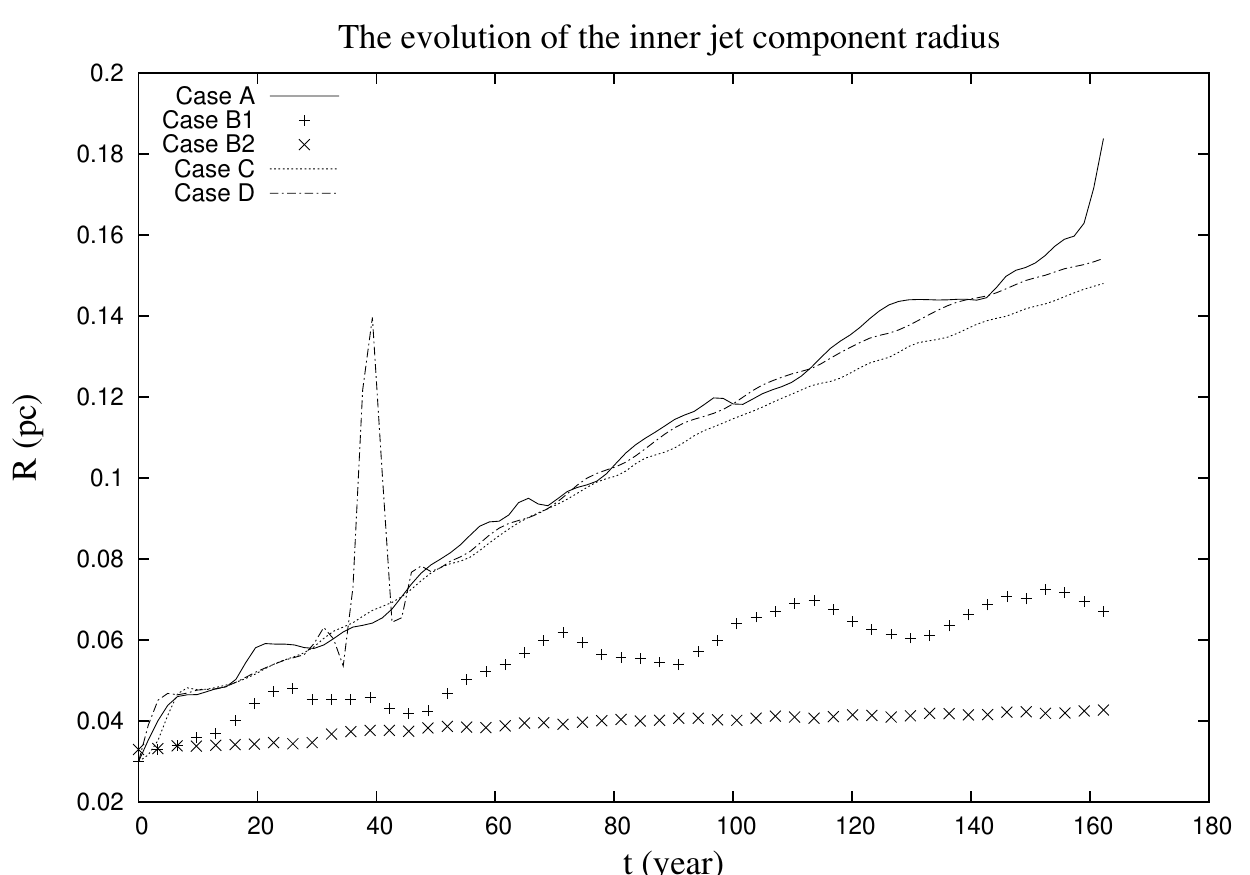}}}
}
\caption{Time evolution of the effective inner jet radius in all cases. For the cases (A), (C), and (D) we distinguish between the inner and outer jet according to the polytropic index and Lorentz factor. 
For the cases (B1) and (B2) we use the same condition for the Lorentz factor and the condition that the inner and 
shear component are magnetized. The effective inner jet radius is $\sqrt{S_{\rm in}/\pi}$ where $S_{\rm in}$ is the surface thus found for the inner jet. The jets (B1, B2) with low inner kinetic energy flux contribution are more stable and remain relativistic and the jets (A, C, D) with high inner kinetic energy flux contribution are unstable and decelerate}\label{Inradius_evolution}
\vspace{-0.1cm}
\end{center}
\end{figure}

\section{Relativistically enhanced Rayleigh-Taylor}\label{lninst}

In cases (A), (C), and (D), a relativistic, Rayleigh-Taylor type instability developed. 
Here, we first describe in a qualitative way how this instability ensues, and then continue to explain it using an approximate linearization.

At the interface between the two outflow components, we mentioned that (linear and then non-linear) 
surface modes develop, due to velocity shear hence of Kelvin-Helmholtz 
type. Fluid of lower inertia (low $\gamma^2 \rho h$) and higher pressure then penetrates the inner jet. Since the inertia in the penetrating spikes is lower 
than in the surrounding inner jet material, the centrifugal force acting on these is weaker. The spikes then expand inwards until the pressure is balanced 
by the centrifugal force. They expand sideways as well, again on the basis of pressure balance and ram pressure arguments.
Due to the big difference in inertia, transfer of angular momentum from the inner jet to these inward penetrating spikes is very efficient. Rather
quickly then, their rotation speed reaches the rotation speed of the inner jet at the initial interface, set by 
$v_{\varphi, \rm in}$. 
During its further inward expansion, this rotation speed remains constant. 
However, the low inertia of the spikes keeps their angular momentum lower than the inner jet matter.

To explain this phenomenon semi-analytically, we confine our attention to a perturbation depending only on the radial direction $R$. 
Let ${\rm d} R=\tilde{v}_{\rm R}{\rm dt}$ represent the radial displacement of the contact surface between the two components during ${\rm dt}$, and $\tilde{v}_{\rm R}$ 
the perturbation speed.  
The momentum equation governing the fluid near our initial equilibrium writes as
\begin{equation}\label{euler}
\left( \gamma^2\,\rho\,h+B_{\rm z}^2 \right)\left[\frac{\partial}{\partial t}+\vec{V}\cdot\nabla\right]\vec{V}
+\nabla p_{\rm total}+\vec{V}\frac{\partial p_{\rm total}}{\partial t}+\ldots =0\,,
\end{equation}
with $ p_{\rm total}=p+\frac{B^2}{2}$.
In this equation we already ignored the lab frame contribution of the charge 
separation, which is justified for rotational flows 
far within the light cylinder, as is the case in our model here~\footnote{Note that we mentioned a jet radius of $0.1 {\rm pc}$ at a distance of $1 {\rm pc}$ for M87. The estimated light cylinder radius for M87 is smaller than $0.1 {\rm pc}$ if a large fraction of the angular momentum (then Poynting flux) extracted from the accretion by the magnetic field is transferred to the stream lines during the acceleration phase by the ideal RMHD mechanisms \citep{Begelman&Li94}. But in our model we assume that the jet is accelerated also by non-ideal RMHD mechanisms with  decay of the toroidal magnetic field by dissipation/reconnection, leading to a fast central spine jet, and overall low rotation profiles (weak transfer of angular momentum from  magnetic fields to stream lines).}
Moreover, the magnetic topology taken here does not involve magnetic pinching, hence magnetic effects only add to total pressure and to the effective inertia, where also $V_{\varphi}<<V_{\rm z}$ is used. 
Then the radial component of Eq.~(\ref{euler}) can be further approximated, if we neglect temporal variation of the total pressure and use $\frac{\partial \tilde{v}_{\rm R}}{\partial \varphi}=0$, to get
\begin{equation}\label{approxeuler}
\left( \gamma^2\,\rho\,h+B_{\rm z}^2 \right)\left[\frac{\partial \tilde{v}_{\rm R}}{\partial t}-\frac{V_{\varphi}^2}{R}\right]+\frac{\partial p_{\rm total}}{\partial R}=0\,.
\end{equation}

According to the initial condition, the inertia  $\gamma^2\,\rho\,h$ of 
the fluid varies slowly in the inner and outer jet. We can thus argue that both inertia and toroidal speed in each fluid element varies slowly 
when it undergoes a small radial displacement $\zeta=R_{\rm in}-R=\int{\tilde{v}_{\zeta} {\rm dt}}=-\int{\tilde{v}_{\rm R} {\rm dt}}$ from the initial interface position $R_{\rm in}$, with $\mid \zeta \mid<<R_{\rm in}$.
Using the main equilibrium balance between centrifugal force and total pressure gradient, the variation of the 
total pressure can be argued to lead to
\begin{equation}\label{lineareuler}
\tilde{p}_{\rm total}=\left(\gamma^2\,\rho\,h+B_{\rm z}^2 \right)\left[\frac{\partial \int{\tilde{v}_{\zeta} {\rm d\zeta}}}{\partial t}+\frac{V_{\varphi}^2\zeta}{R_{\rm in}}\right]\,.
\end{equation}
To get an approximate dispersion relation, we assume that to first order, the perturbation speed is potential, 
i.e. $\tilde{v}_{\zeta}=\left(\nabla \Psi\right)_\zeta$ and that we have $\Psi\propto \exp(\lambda\,t-k\,\mid \zeta\mid)$. 
Noting that the displacement $\zeta=\int{\tilde{v}_{\zeta} {\rm dt}}$ must be identical for inner/outer regions, and using
total pressure and displacement continuity arguments, we
then get the essential proportionality relation between the instability growth rate and wave number
\begin{equation}\label{disprel}
\lambda^2 \propto k\,\left[\left(\gamma^2\,\rho h+B_{\rm z}^2\right)_{\rm in} -\left(\gamma^2\,\rho h+B_{\rm z}^2\right)_{\rm out}\right] \,.
\end{equation}
This approximative dispersion relation indicates that for two-component stability, we need $\lambda^2<0$ requiring that $\left(\gamma^2\,\rho h+B_{\rm z}^2\right)_{\rm out}>\left(\gamma^2\,\rho h+B_{\rm z}^2\right)_{\rm in}$. 
This confirms that the interface between the two rotating components is stable against a centrifugally driven, 
relativistic Rayleigh-Taylor instability, when the effective inertia of the outer component is higher than the 
effective inertia of the inner component. This important result correlates well with the results of all 
simulations in section~\ref{Simulations} as well as in our previous publication~\citep{Meliani&Keppen07b}. 
In fact, this equation explains why cases (A), (C), (D) are relativistically  Rayleigh-Taylor unstable 
since we find  typical contrasts during the evolution in (A) $\left(\gamma^2\,\rho h+B_{\rm z}^2\right)_{\rm out}\approx 0.75 {\left(\gamma^2\,\rho h+B_{\rm z}^2\right)_{\rm in}}$, in (C) and (D) $\left(\gamma^2\,\rho h+B_{\rm z}^2\right)_{\rm out}\approx 0.1 {\left(\gamma^2\,\rho h+B_{\rm z}^2\right)_{\rm in}}$, whereas cases (B1) and (B2) are relativistically Rayleigh-Taylor stable since 
$\left(\gamma^2\,\rho h+B_{\rm z}^2\right)_{\rm out}\sim 10 \left(\gamma^2\,\rho h+B_{\rm z}^2\right)_{\rm in}$ in (B1) and  
${\left(\gamma^2\,\rho h+B_{\rm z}^2\right)_{\rm out}}\sim 20{\left(\gamma^2\,\rho h+B_{\rm z}^2\right)_{\rm in}}$ in (B2). 
 Initial effective inertia ratios derived from the initial conditions are mentioned in Table~\ref{table:1}, and it is clear that we can set up stable versus unstable cases by varying the contrast of effective inertia between inner versus outer jet. This also works in pure hydro, and indeed the pure hydro case from~\cite{Meliani&Keppen07b} does not suffer from this newly discovered instability. For the purely poloidal field configurations studied here, magnetic field effects are prominent in total pressure and effective inertia alone.
Note finally that it is a truly relativistic effect, since the same argument in classical MHD just involves the
density difference between outer versus inner jet. Under the conditions of a light inner jet with heavy outer jet
taken here, a classical variant of relation~(\ref{disprel}) involves only $\rho_{\rm out}-\rho_{\rm in}$ and
would predict stability.

\begin{figure}[h]
\begin{center}
\FIG{
{\resizebox{0.95\columnwidth}{6cm}{\includegraphics{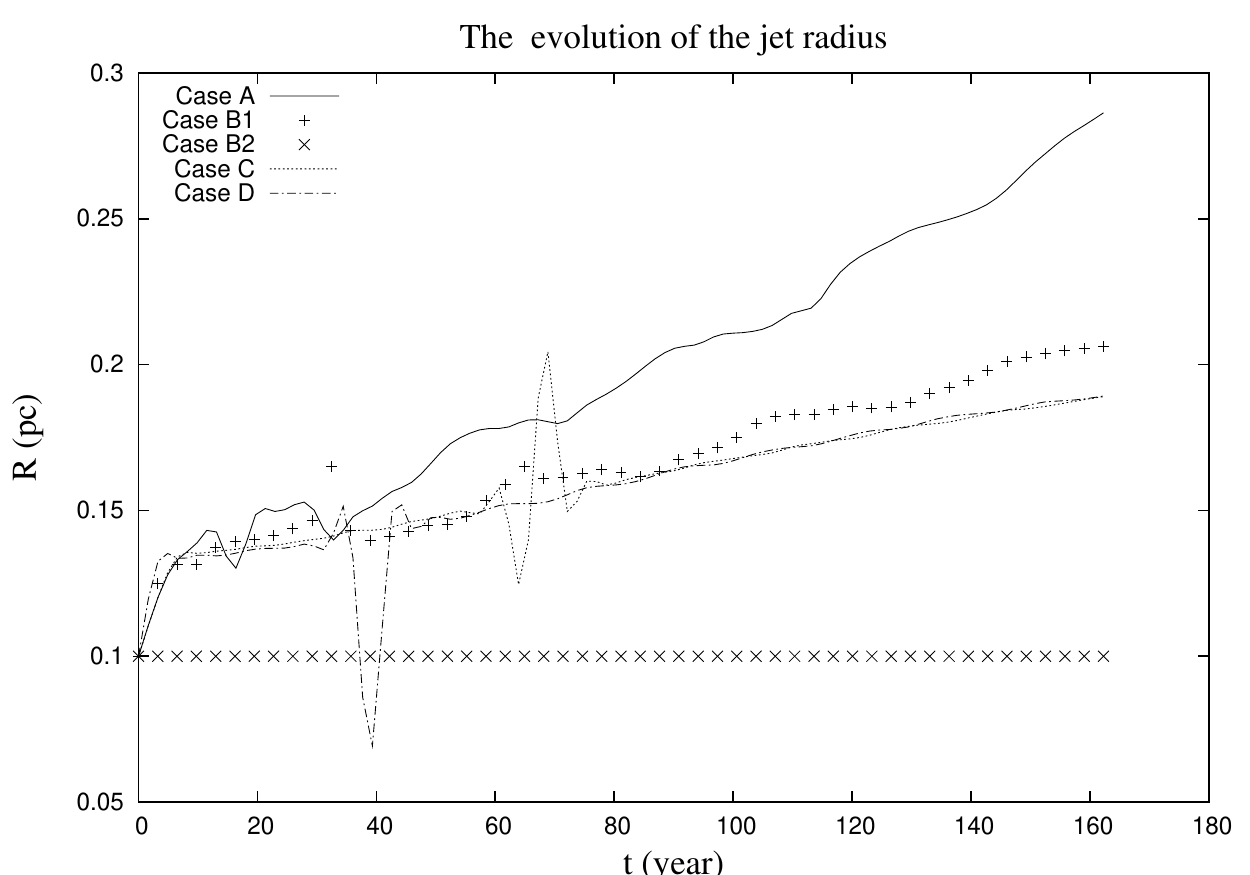}}}
}
\caption{Time evolution of the effective jet radius in all cases. The effective jet radius is $\sqrt{S_{\rm jet}/\pi}$ where $S_{\rm jet}$ is the surface occupied by the full (inner+outer) jet.  The jets (A, C, D) with high inner kinetic energy flux contribution  decollimate.}\label{Radius_evolution}
\end{center}
\end{figure}

\section{Discussion and conclusions}

We examined five configurations of magnetized two-component jets. All share the same density ratio between inner and outer component
and identical rotation profiles. The magnetic and thermal pressure configuration in each model differs, though.
This in turn translates to different distributions of the (total, fixed) kinetic energy flux over the inner and outer jet component. 
In fact, in case (A) the kinetic energy flux in the inner component is about $10\%$, while it is $0.7\%$ for case (B1) and $0.5\%$ in case (B2), and around $38\%$ for cases (C) 
and (D). For the outer component, we thus have $90\%$ in case (A), $99.3\%$ in case (B1) and $99.5\%$ in 
case (B2), and only $62\%$ for cases (C) and (D). 
The most important difference between the two model categories is then: cases (A, C, D) 
have an inner jet component with higher inertia $\gamma^2\,\rho\,h+B_{\rm z}^2$ than their outer jet 
component, while cases (B1) and (B2) have an inner jet component inertia which ends up lower 
than in the outer jet component. From the detailed analysis of the simulations, as well as from the approximate stability analysis, this criterion
distinguishes between cases where relativistically enhanced Rayleigh-Taylor modes ultimately dominate the evolution, leading to complete mixing
of both components and inner jet deceleration. 
This is quantified most clearly by showing the time evolution of the mean Lorentz factor over the inner jet region
for all cases in Fig.~\ref{Lorentzfactor_evolution}. This requires a clear criterion to distinguish
inner versus outer jets in the turbulent evolutions.
In cases (A), (C) and (D), we locate the outer jet component as having a Lorentz factor $2.5<\gamma<3.5$ and 
effective polytropic index $\Gamma_{\rm eff}>3/2$. The inner component is the region defined by a 
Lorentz factor $\gamma\ge 3.5$ and effective polytropic index $\Gamma_{\rm eff}\le 3/2$. 
In the cases (B1) and (B2), the effective polytropic index in the inner and outer jet can be locally of the 
same order there, hence we use that the inner jet is magnetized. In the cases (B1) and (B2), 
the inner component jet and shear region are not totally mixed during the evolution, since the inner component is 
compressed and has higher magnetization and Lorentz factor. Thus to distinguish the inner component jet we put a 
condition on magnetic field strength $B_{\rm z}>B_{\rm z, initial}/2$, where $B_{\rm z, initial}$ is the magnetic 
strength assumed initially. Under these precise quantifications of inner/outer jet regions, Fig.~\ref{Lorentzfactor_evolution} demonstrates clearly how stable cases (for the relativistically enhanced Rayleigh-Taylor modes)
remain at high speed, while unstable cases decelerate. Using the same means to distinguish inner versus outer jet regions at all times, we can quantify the inner jet radius for all cases, as well as the total jet radius for all cases. These are shown in Figs.~\ref{Inradius_evolution}-\ref{Radius_evolution}, and quantify the 
decollimation effects discussed in Section~\ref{Simulations}.

During the entire evolutions, the toroidal and radial speeds remain weak as we have typically a maximal $V_{R}<0.01$ and $V_{\varphi}<0.04$. 
This means that the contribution of the laboratory frame charge separation force $\rho_{e}{\vec{E}}$ to the Lorentz force is negligible at all times.
Under these conditions, in both radial and toroidal directions, the contribution of magnetic energy to fluid inertia is $1/\gamma^2$ weaker than the
contribution of the magnetic pressure to the total pressure. This explains why cases (B1) and (B2) are then more stable than the other cases. Despite the fact 
that case (B2) has similar axial total pressure than  other cases (A, C, D), the low contribution of thermal energy to total pressure 
makes the effective inertia ratio between the inner and outer jet low. This two-component jet is then stable against the relativistically enhanced
Rayleigh-Taylor instability. Extraction of angular momentum and energy from the inner jet to the shear shell is less efficient in cases (B1) and (B2) than in all
other cases. As a result, in cases (B1) and (B2) the inner relativistic jet with high Lorentz factor $\gamma_{\rm in}\sim 20$  persists. In the other cases, 
the kinetic energy flux in the inner jet was initially relatively high, making them unstable and leading to deceleration to Lorentz factors 
around $8$.

We investigated the stability of two-component jets  beyond the launching region, where both components are rotating differently and a clear
two-component structure exists. We initialized this model in accord with magneto-centrifugal models for jet generation, also using the observed
analogy between radio source jets and two-component jets in young stellar objects, where the rotation within the jet can actually be observed.
We performed five very high resolution simulations of magnetized two-component jets with various magnetization and kinetic energy flux stratifications.
The two-component jets with a low inner kinetic energy flux contribution are more stable and remain relativistic for long distances, whereas jets with a 
highly contributing inner jet to the total jet kinetic energy flux, are subject to a relativistic Rayleigh-Taylor type instability. This instability 
turns out to be very efficient to decollimate and decelerate the inner jet. Jets that are subject to this instability become turbulent after propagating
for a distance of about $30 {\rm pc}$. 

This new result on two-component jet models is important because it can explain the classification of radio 
sources in Fanaroff-Riley I/II categories according the energy stratification of the inner jet.
This ultimately relates to the jet launch region and the properties of the inner accretion disk. 
In fact, by analogy between the FRI/FRII classification  and the results of our model, an FRI jet would
correspond to a two-component jet with a high energy flux contribution from the inner jet, 
whereas the FRII jet corresponds to relatively low energy fluxes in the inner jet.
The model we propose here to explain the FRI/FRII dichotomy is different from the model we proposed 
earlier~\citep{Melianietal08} where the transition occurs due to external density stratification. That model 
explains the group of peculiar ``HYbrid MOrphology Radio Source" (HYMORS) \citep{GopalKrishna&Wiita00} 
which appear to have an FR II type on one side and an FR I type diffuse radio lobe on the other side of the 
active nucleus. Since the launch conditions on each side are presumably similar in these kind of radio sources, 
the different Fanaroff-Riley morphologies on either side must be attributed to the change in the properties of 
the ambient media, as shown convincingly in~\cite{Melianietal08}. The results of the present paper
nicely complement these earlier findings with a quantifiable role of the central engine contribution.
 
We currently continue this study of the interaction between two component jets in full 3D. We thereby intend to 
explore the relative influence of azimuthal versus 
longitudinal instabilities for realistic multi-component jets. Another extension is to allow for aximuthal
magnetic fields, in accord with the initial profiles as given generally in this paper.  It then rremains to be shown that (1) the newly discovered instability persists in 3D hydro and magnetohydrodynamic configurations, where the potential role of axial mode development is incorporated, and introduces helical jet axis displacements; (2) how the instability gets modified (stabilized or destabilized) by the inclusion of toroidal field components, first in 2.5D neglecting helical axis displacements, and consecutively in 3D, where also current-driven kinks may occur.

\begin{acknowledgements}
We acknowledge financial support from the FWO, grant G.027708, and from the K.U.Leuven GOA/09/009.
Part of the computations made use of the High Performance Computing VIC cluster at K.U.Leuven.
Z. Meliani  acknowledge financial support from HPC Europa (project number: 228398). Visualization was performed using Paraview, see {\tt www.paraview.org}.
\end{acknowledgements}

\bibliographystyle{aa}

\end{document}